\documentclass[useAMS,usenatbib]{mn2e}

\usepackage{natbib}
\usepackage{graphicx}
\usepackage{listings}
\usepackage[intlimits]{amsmath}
\usepackage{txfonts}
\usepackage{bm}
\usepackage{amssymb}
\usepackage{hyperref}
\usepackage{amsmath}
\usepackage{tabularx} 
\usepackage{color}

\newcommand{\beq}{\begin{equation}}
\newcommand{\eeq}{\end{equation}}

\def\be{\begin{equation}}
\def\ee{\end{equation}}
\def\ba{\begin{eqnarray}}
\def\ea{\end{eqnarray}}
\def\go{\mathrel{\raise.3ex\hbox{$>$}\mkern-14mu
             \lower0.6ex\hbox{$\sim$}}}
\def\lo{\mathrel{\raise.3ex\hbox{$<$}\mkern-14mu
             \lower0.6ex\hbox{$\sim$}}}

\def\apj{Astrophys. J.}
\def\apjl{Astrophys. J. Lett.}
\def\apjs{Astrophys. J. Supp. Ser. }

\def\aap{Astron. Astrophys. }

\def\araa{Ann.\ Rev. Astron. Astroph. }

\def\mnras{Mon. Not. Roy. Astron. Soc. }

\def\nat{Nature}

\def\pre{Phys. Rev. E.}

\title[Powering CCOs with a Tangled Magnetic Field]{Powering Central Compact Objects with a Tangled Crustal Magnetic Field}

\author[K.N. Gourgouliatos, R. Hollerbach and A.P. Igoshev]{{Konstantinos N. Gourgouliatos $^{1}$}\thanks{kngourg@upatras.gr}, {Rainer Hollerbach  $^{2}$} and {Andrei P. Igoshev$^{2}$}\vspace{0.4cm}\\
\parbox{\textwidth}{$^{1}$University of Patras, Department of Physics, 26504, Patras, Greece }\\
\parbox{\textwidth}{$^{2}$University of Leeds, Department of Applied Mathematics, LS2 9JT, Leeds, UK} }

\begin{document}

\date{Accepted -. Received -; in original form -}
\pagerange{\pageref{firstpage}--\pageref{lastpage}} \pubyear{-}
\maketitle

\label{firstpage}

\begin{abstract}Central Compact Objects (CCOs) are X-ray sources with luminosity ranging between $10^{32}$-$10^{34}$ erg~s$^{-1}$, located at the centres of supernova remnants. Some of them have been confirmed to be neutron stars. Timing observations have allowed the estimation of their dipole magnetic field, placing them in the range $\sim10^{10}$-$10^{11}$ G. The decay of their weak dipole fields, mediated by the Hall effect and Ohmic dissipation, cannot provide sufficient thermal energy to power their X-ray luminosity, as opposed to magnetars whose X-ray luminosities are comparable. Motivated by the question of producing high X-ray power through magnetic field decay while maintaining a weak dipole field, we explore the evolution of a crustal magnetic field that does not consist of an ordered axisymmetric structure, but rather comprises a tangled configuration. This can be the outcome of a non-self-excited dynamo, buried inside the crust by fallback material following the supernova explosion. We find that such initial conditions lead to the emergence of the magnetic field from the surface of the star and the formation of a dipolar magnetic field component. An internal tangled magnetic field  of the order of $10^{14}$ G can provide sufficient Ohmic heating to the crust and power CCOs, while the dipole field it forms is approximately $10^{10}$ G, as observed in CCOs.
\end{abstract}


\section{Introduction}

Simple magnetic flux conservation in a neutron star progenitor can provide a magnetic field up to $\sim 10^{12}$ G, assuming a moderately magnetised progenitor. While this value is sufficient for most rotation-powered pulsars, it falls below the spin-down dipole field of the strongly magnetised ones and the magnetar population. Thus, some process amplifying the strength of the magnetic field  needs to take place during the formation of neutron stars. For instance, if the collapsing neutron star rotates differentially, a seed dipolar field will create an azimuthal field \citep{Spruit:2008}. More likely, dynamo action may take place during the formation of the neutron star, amplifying the magnetic field strength \citep{Thompson:1993,Reboul:2019}. While dynamo mechanisms have been assumed in order to generate a dipole field,  they typically involve non-axisymmetric configurations. Therefore, even if a large scale dipole field forms, most likely the global field will also have non-axisymmetric features. 

During the very early stages of neutron star evolution and prior to the freezing of the neutron star crust, it is possible to have dynamo action amplifying the magnetic field strength. While this stage lasts only for a few seconds, it corresponds to several turnover times and is sufficiently long for the system to relax to some equilibrium, or for the exponential growth to saturate \citep{Rembiasz:2016, Raynaud:2020}. An efficient dynamo mechanism requires rapid rotation of the collapsing object \citep{Bonanno:2006}; if this is not the case, it is likely that while loops of magnetic field form through convective activity, the system never reaches the stage where a strong dipolar magnetic field appears. This mechanism has been studied in \cite{Thompson:2001} where a stochastic dynamo operates on the material accreted onto supernova cores and leads to the formation of convective shells containing magnetic field of $10^{14}$ G. This evolutionary path leads to a newborn neutron star where the initial magnetic field is strong but highly non-dipolar. Following the dynamo stage and while the star is still fluid (prior to crust crystallisation), the outcome of dynamical relaxation may lead to various outcomes. One possibility is that the magnetic field relaxes to a twisted-torus configuration \citep{Braithwaite:2004, Braithwaite:2006}, where the magnetic field is predominantly dipolar and contains a toroidal magnetic field \citep{Ciolfi:2013}, but it can also adopt equilibria in the form of highly tangled structures, with non-axisymmetric multipolar configurations, which are stable despite their complexity \citep{Braithwaite:2008}. 

From an observational perspective, the structure of the magnetic field in neutron stars cannot be determined in fine detail. The field strength of a neutron star usually refers to the dipole component deduced by some spin-down model, assuming a given moment of inertia. The canonical pulsar model postulates an oblique dipole rotating in vacuum and spinning-down due to electromagnetic dipole radiation, with its moment of inertia being that of a sphere with uniform density. In some exceptional cases of  strongly magnetised neutron stars the small scale magnetic field has been deduced \citep{Guver:2011, Tiengo:2013}. These observations rely on phase dependent absorption features where the X-ray emission originating from the surface is partially absorbed by particles trapped on magnetic field lines of the small scale magnetic field. 

Such complex magnetic fields may not be present only to magnetars, but also to an other family of neutron stars with puzzling behaviour, the so-called CCOs. There are ten such sources (confirmed and candidate) \citep{DeLuca:2017}, all detected in the soft X-ray part of the spectrum, which are located at the centres of supernova remnants and are believed to be radio quiet, isolated, young neutron stars. Three of  them have been timed with periods of $105$ ms, $112$ ms and  $424$ ms \citep{Zavlin:2000, Gotthelf:2005, Gotthelf:2009}, with the deduced spin-down dipole field corresponding to $\lesssim 10^{11}$ G, while their characteristic age is inconsistent by several orders of magnitude with the age of the host supernova remnant, which is on the order of a few kyrs. Considering their bolometric luminosities, several CCOs have $L_x>10^{33}$ erg s$^{-1}$, which places them closer to the magnetar family \citep{Olausen:2014} rather than to rotation-powered pulsars, whose magnetic fields have similar strengths. Their high X-ray luminosities cannot be attributed to residual cooling of the proto-neutron star, since this takes place rapidly within the first few $100$ years \citep{Yakovlev:2004}. Magnetic field decay by Ohmic dissipation could provide the required thermal energy, but this would require magnetar-level fields \citep{Halpern:2010}. While a strong magnetic field resolves the question of the origin of their X-ray luminosity, it has some critical side-effects: an internal field  of $10^{15}$ G  may lead to crustquakes and activity similar to that of magnetars, which is not typical of CCOs\footnote{We note that source 1E 161348-5055, which is located at the centre of a supernova remnant, has given outbursts, which is typical of magnetars. Quite possibly, this object is actually a magnetar that has undergone a phase of accretion burying its magnetic field and slowing down its rotation rate to extreme levels \citep{Rea:2016}.}. Thus, an interesting question arises: do field configurations exist that are strong enough to decay and provide sufficient heat, but not so strong that they would fracture the crust, which would yield more typical magnetar behaviour with bursts and flares rather than CCO-like behaviour. Such a magnetic field cannot have a dipole component, otherwise CCOs would appear more strongly magnetised. Moreover, the presence of absorption features in the spectra of CCOs favours the existence of a predominantly non-dipolar field in CCOs with a complex configuration \citep{Sanwal:2002, DeLuca:2004, Gotthelf:2013}. A possible resolution of this puzzle could be a highly tangled magnetic field of moderate strength. Such a field would be supported by strong currents and produce sufficient Ohmic heating, but the stresses the field exerts on the crust would not be strong enough to make it yield and power bursts.  
 
 The magnetic field of neutron stars can still evolve at a much longer timescale of kyrs, compared to the dynamo timescales mentioned earlier. This happens despite the fact that the crust has reached a dynamical equilibrium, but still the mechanisms of Hall drift, Ohmic decay and ambipolar diffusion operate and drive magnetic field evolution \citep{Goldreich:1992}. To follow the long term changes of a tangled magnetic field one needs to simulate the dominant effects driving its evolution. Significant progress has been made in the theoretical modelling of the crustal magnetic field evolution in neutron stars, with numerous studies exploring in detail the evolution of the Hall effect and Ohmic dissipation. Analytical and numerical approaches in various geometries have demonstrated the highly non-trivial paths magnetic field evolution can take \citep{Hollerbach:2002, Hollerbach:2004, Cumming:2004, Vainshtein:2000, Reisenegger:2007, Pons:2007, Pons:2009, Vigano:2012, Vigano:2013, Gourgouliatos:2014a, Pons:2019}. In addition to the Hall-MHD evolution of the magnetic field, further effects have been studied, such as ambipolar diffusion \citep{Passamonti:2017a, Castillo:2017}, superconductivity \citep{Passamonti:2017b}, the elastic response of the crust, its failure and the consequent plastic flows \citep{Li:2016, Bransgrove:2018, Lander:2016, Thompson:2017, Lander:2019}. 

A key input for the numerical studies is the initial magnetic field structure. The early stages of the magnetic field evolution are strongly determined by the details of the magnetic field structure and in particular the ratio between the poloidal and toroidal field. Axisymmetric initial conditions dominated by the poloidal magnetic field generate configurations where the field maintains its initial axisymmetric structure \citep{Wood:2015, Gourgouliatos:2016b}. In contrast, if the initial state is dominated by the toroidal component of the magnetic field, the evolution is susceptible to instabilities \citep{Rheinhardt:2002, Pons:2010, Wood:2014, Gourgouliatos:2015b, Gourgouliatos:2016a} which then generate non-axisymmetric features \citep{Gourgouliatos:2018, Gourgouliatos:2019}. Apart from axisymmetric initial conditions, it is also possible that the initial magnetic field structure is dominated by intermediate scale features, rather than a strong dipolar magnetic field. Initial conditions consisting of higher order poloidal multipoles have been explored in axisymmetric simulations \citep{Igoshev:2016}, however such tangled magnetic fields have not yet been studied in detail in 3-D crust shell simulations. 

Motivated by the puzzle of CCOs, we explore the magnetic field evolution in the crust with initial conditions that represent the endpoint of a stochastic dynamo. We populate the crust with a magnetic field consisting of loops whose radius is comparable to the thickness of the  crust, while setting the initial value of the dipolar component either zero, or a few orders of magnitude less than the average field. We then follow the magnetic field evolution by simulating the Hall effect and Ohmic dissipation, and we extract observable quantities. 

The structure of the paper is as follows. We set up the mathematical equations and numerical methods applied to the problem in section \ref{MATH}, and present the results of the numerical runs in section \ref{RESULTS}. Section \ref{DISCUSSION} discusses the implications of these results to CCOs and neutron stars in general. We conclude in section \ref{CONCLUSIONS}.

\section{Mathematical and numerical setup}
\label{MATH}

The crust of a neutron star can be approximated by a crystal lattice consisting of ions and free electrons, with the latter carrying the electric current. This allows a single fluid approximation, the so-called electron or Hall MHD. The evolution of the magnetic field in this system is described by the following induction equation \citep{Goldreich:1992}:
\begin{equation} 
\frac{\partial\vec B}{\partial t}= - {\nabla \times}\left(\eta 
\nabla \times \vec{B}  + \frac{c}{4 \pi e n_e} (\nabla \times \vec{B} ) \times \vec B  \right) \,,
\label{Hallind} 
\end{equation} 
where ${\vec B}$ is the magnetic field, $\eta=c^2/(4\pi \sigma)$ is the magnetic diffusivity, $\sigma$ is the electrical conductivity, $e$ is the elementary charge, $n_e$ is the electron number density, and $c$ is the speed of light. To simulate the magnetic field evolution we numerically integrate the above equation in a spherical shell representing the crust. We use a suitably modified version of the PARODY 3-D MHD code \citep{Dormy:1998, Aubert:2008, Wood:2015}, employing Crank-Nicolson and Adams-Bashforth numerical schemes.  Vacuum boundary conditions are imposed at the exterior of the star, and superconductor boundary conditions at the base of the crust, preventing the magnetic field from penetrating into the core. The electron number density profile is approximated by $n_e =2.5\times 10^{34} [(1.0463r_{NS} - r)/0.0463r_{NS}]^4$~cm$^{-3}$, with $r_{NS}$ being the neutron star radius, set to $10$ km. The crust is taken to be the outer $0.1$ of the stellar radius. The electrical conductivity is   $\sigma= 1.8 \times 10^{23}[n_{e}/(2.5\times 10^{34})]^{2/3}$s$^{-1}$, ranging from $1.8 \times 10^{23}$ s$^{-1}$ at the top of the crust to $3.8 \times 10^{24}$ s$^{-1}$ at the base. We use a resolution of 144 radial points, and maximum $\ell=m=80$ in the spherical harmonic decomposition. 

The initial conditions consist of a magnetic field containing several multipoles in the range $10\leq \ell \leq 20$ and $m \leq \ell$ with random phases. The field's intensity peaks at the centre of the crust and decreases rapidly beyond the point where the electron number density is $n_e=5\times 10^{34}$~cm$^{-3}$, corresponding approximately to the neutron drip point and is about 80 m below the conventional neutron star surface of our simulation. This accounts for the burying of the magnetic field that has been proposed to happen shortly after the formation of CCOs by fall-back supernova material.  

We have simulated seven models varying the magnetic energy in the tangled part versus the dipole component. The properties of the models are shown in Table \ref{TAB:1}. We run the models until the age of the neutron star is at least $50$ kyrs, which exceeds the age of the associated supernova remnants hosting the CCO. In models 4 and 5 we run the simulations for 1.5 Gyr, to see the possible outcome of a CCO. 

\begin{table}
\centering
\caption{The seven numerical models, with columns from left to right having the name of the model and the initial magnetic energy, dipole field, and average crustal field. }
\label{TAB:1}
\begin{tabular}{cccc}
\hline
Name	& $E_{tot,0}$ (erg)	&$B_{dip,0}$ (G)& $\bar{B}_0$ (G) \\
\hline
Model 1 & $2.5\times 10^{45} $& $0 $& $2\times 10^{14}$\\
Model 2 & $2.5\times 10^{45} $& $10^{10}$& $2\times 10^{14} $\\
Model 3 & $2.5\times 10^{45} $& $10^{11} $& $2\times 10^{14}$ \\
Model 4 & $2.5\times 10^{47} $& $10^{10} $& $2\times 10^{15} $\\
Model 5 & $2.5\times 10^{47} $& $10^{12} $& $2\times10^{15}$ \\
Model 6 & $4\times 10^{48} $& $10^{10}$& $10^{16}$\\
Model 7 & $4\times 10^{48} $ & $10^{11}$& $10^{16}$\\
\hline
\end{tabular}
\end{table}

\section{Results}
\label{RESULTS}

In all models the magnetic field emerges from the surface, due to the effect of Hall drift, at a timescale that depends on the strength of the tangled magnetic field. Taking the shell average, in 1, 2 and 3 it takes more than 2000 years for the field to emerge from the neutron drip point surface, in models 4 and 5 about 500 years, and in models 6 and 7 about 100 years.  However, because of the anisotropy, regions of strong magnetic field appear on the surface at about half the time it for the average crustal field to increase at the surface.

\begin{figure*}
\centering
{\bf a}\includegraphics[width=3.6cm]{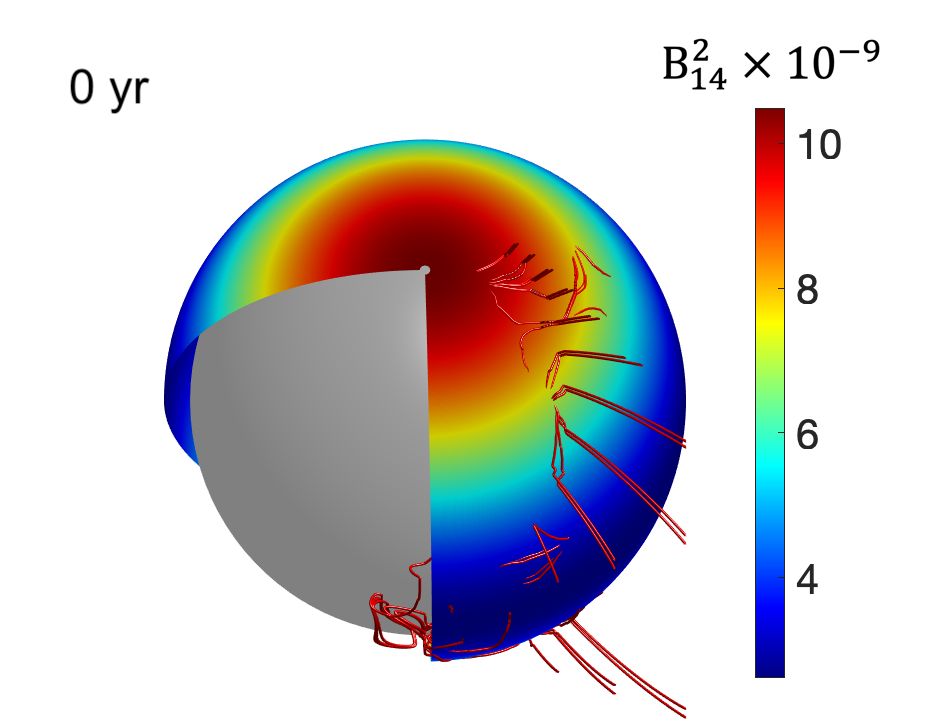}
{\bf b}\includegraphics[width=3.6cm]{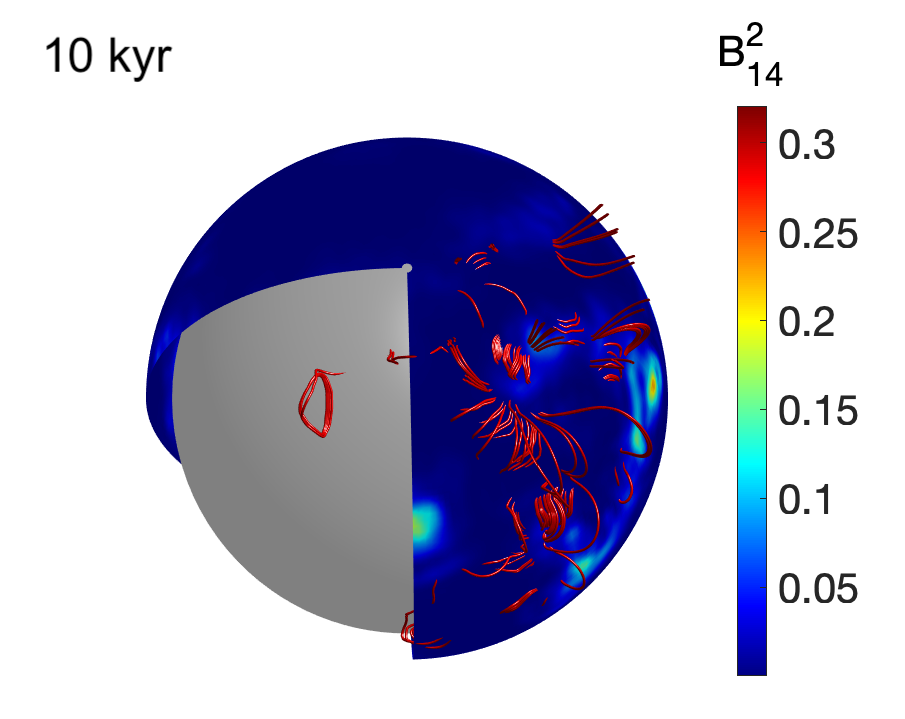}
{\bf c}\includegraphics[width=3.6cm]{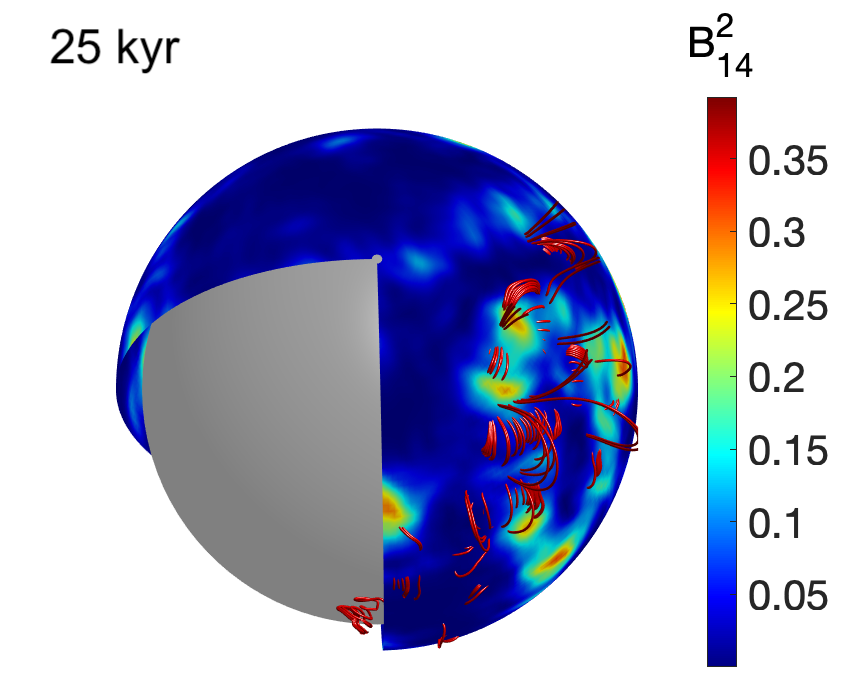}
{\bf d}\includegraphics[width=3.6cm]{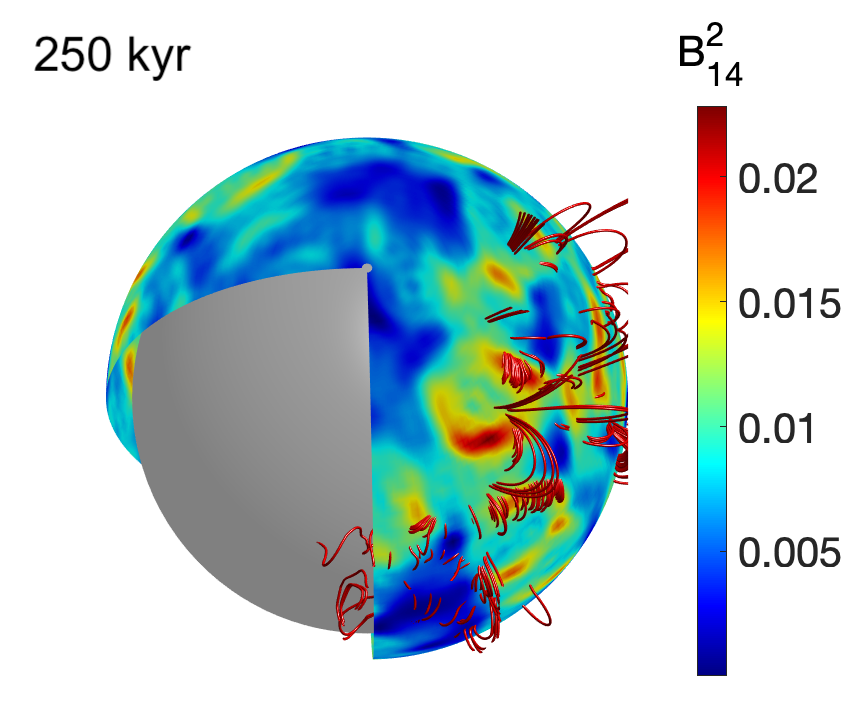}
\caption{The magnetic field structure for Model 2. The colour-plot of the surface shows the magnetic field squared and the field lines are shown in red.}
\label{Fig:1}
\end{figure*}

\begin{figure*}
\centering
{\bf a}\includegraphics[width=3.6cm]{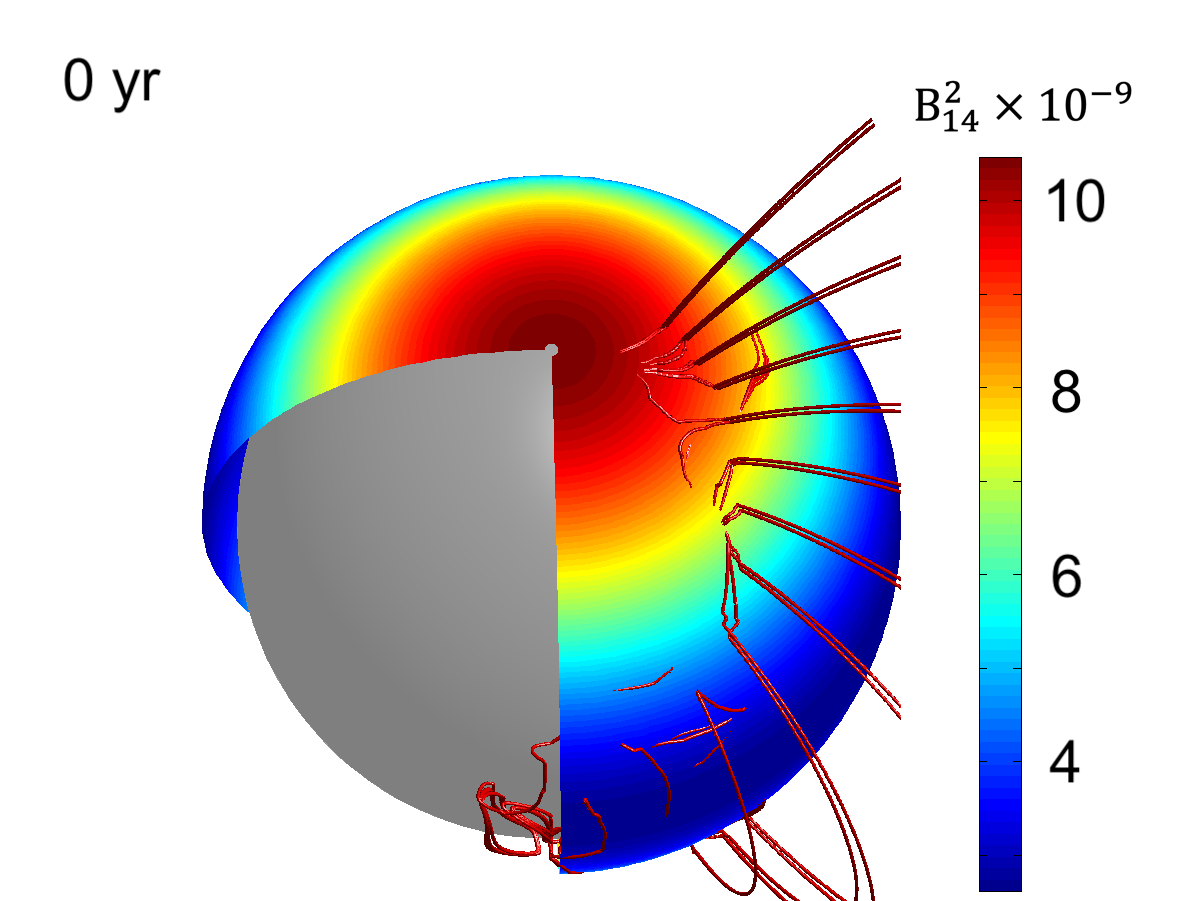}
{\bf b}\includegraphics[width=3.6cm]{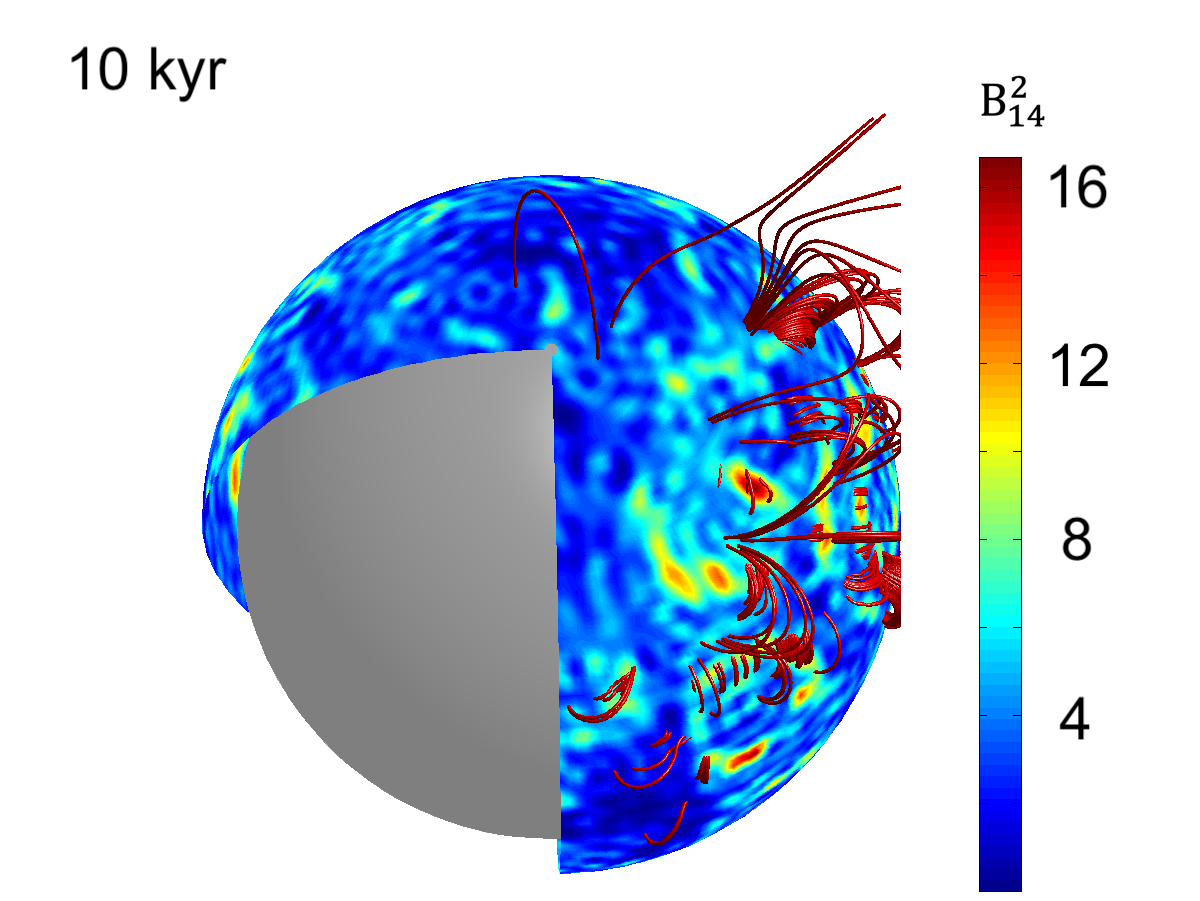}
{\bf c}\includegraphics[width=3.6cm]{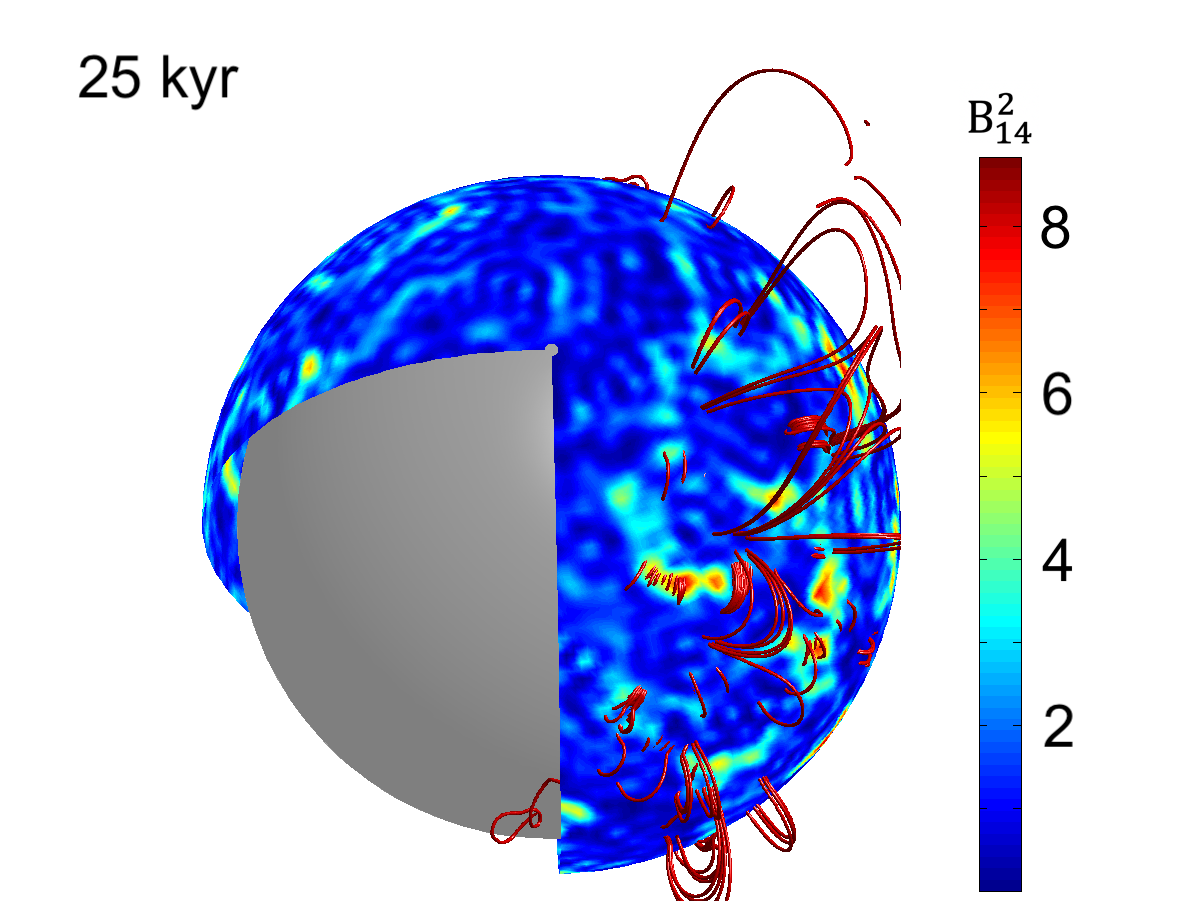}
{\bf d}\includegraphics[width=3.6cm]{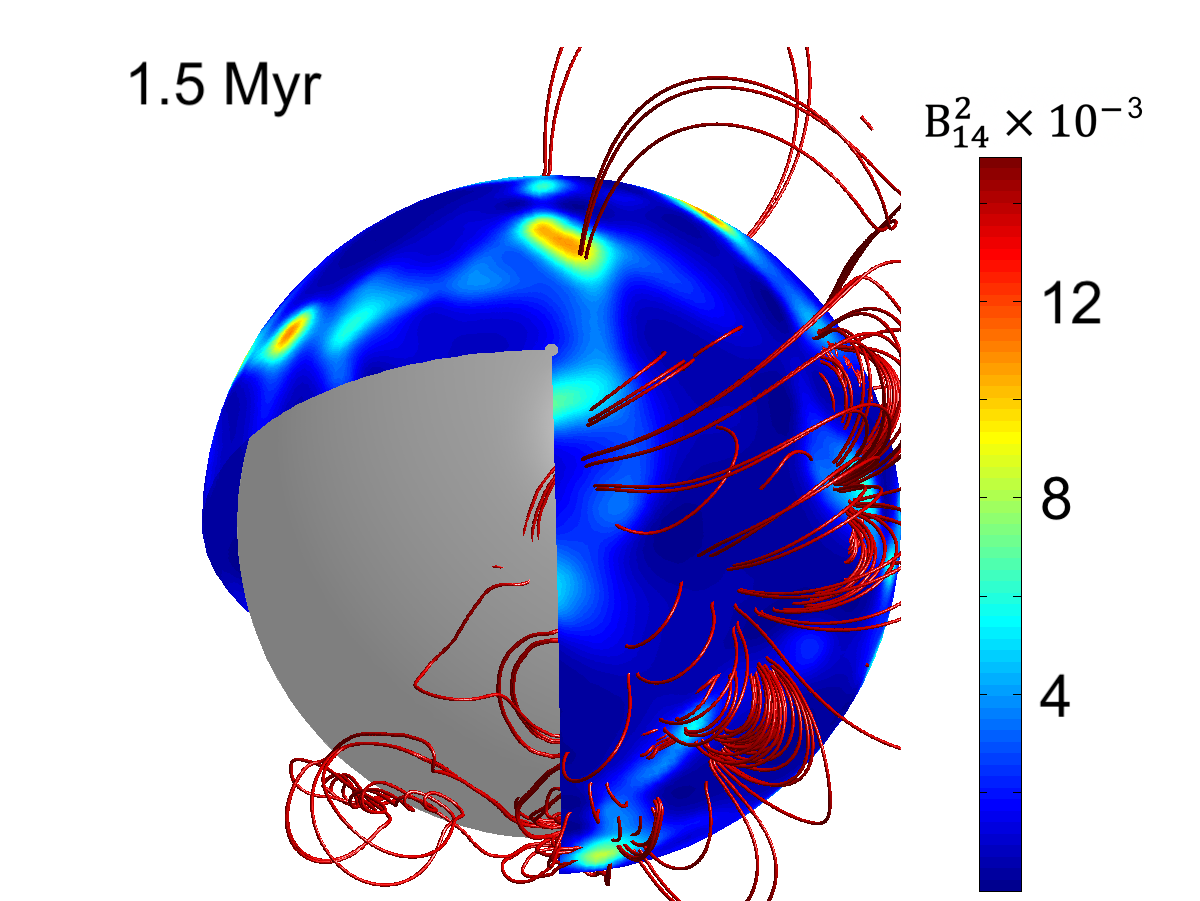}
\caption{The magnetic field structure for Model 4, as in {\bf Fig.~\ref{Fig:1}}.}
\label{Fig:2}
\end{figure*}

\begin{figure*}
\centering
{\bf a}\includegraphics[width=3.6cm]{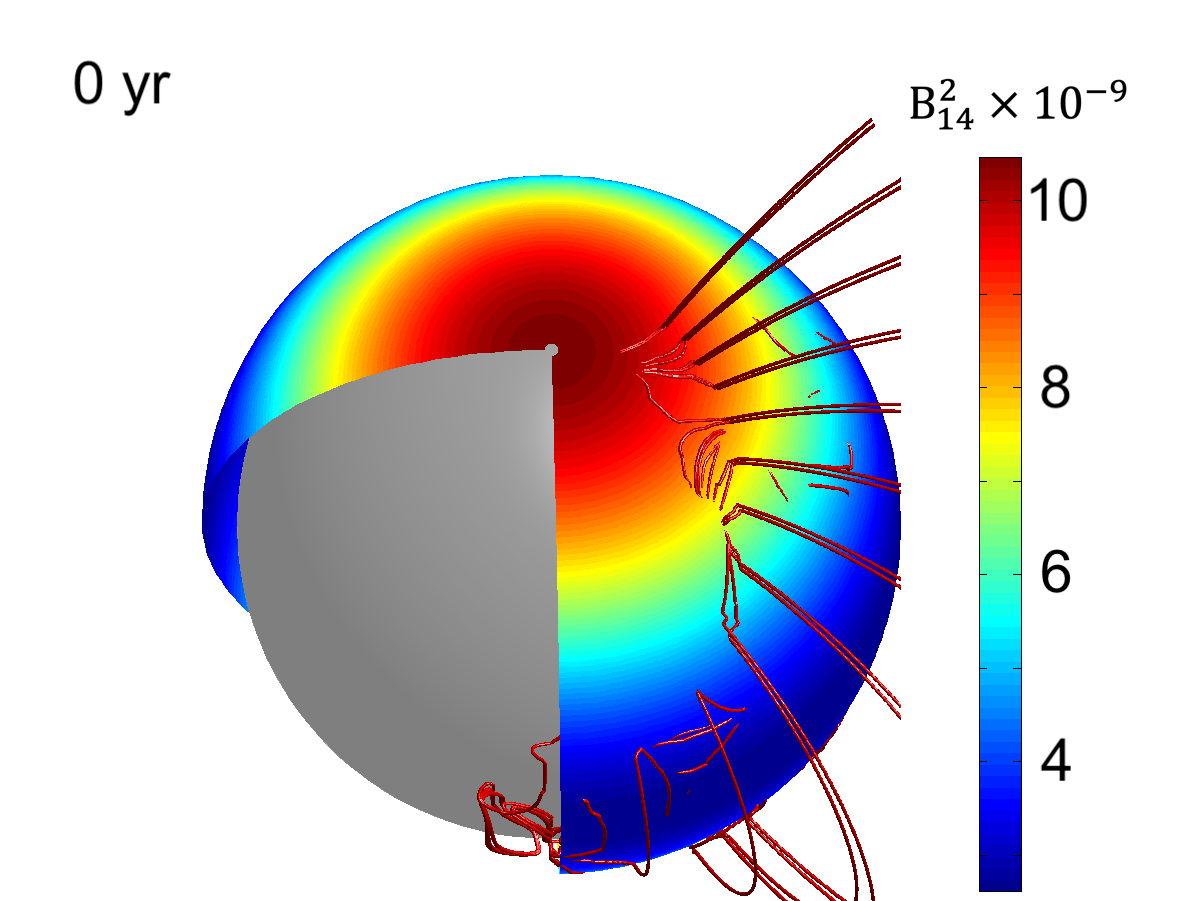}
{\bf b}\includegraphics[width=3.6cm]{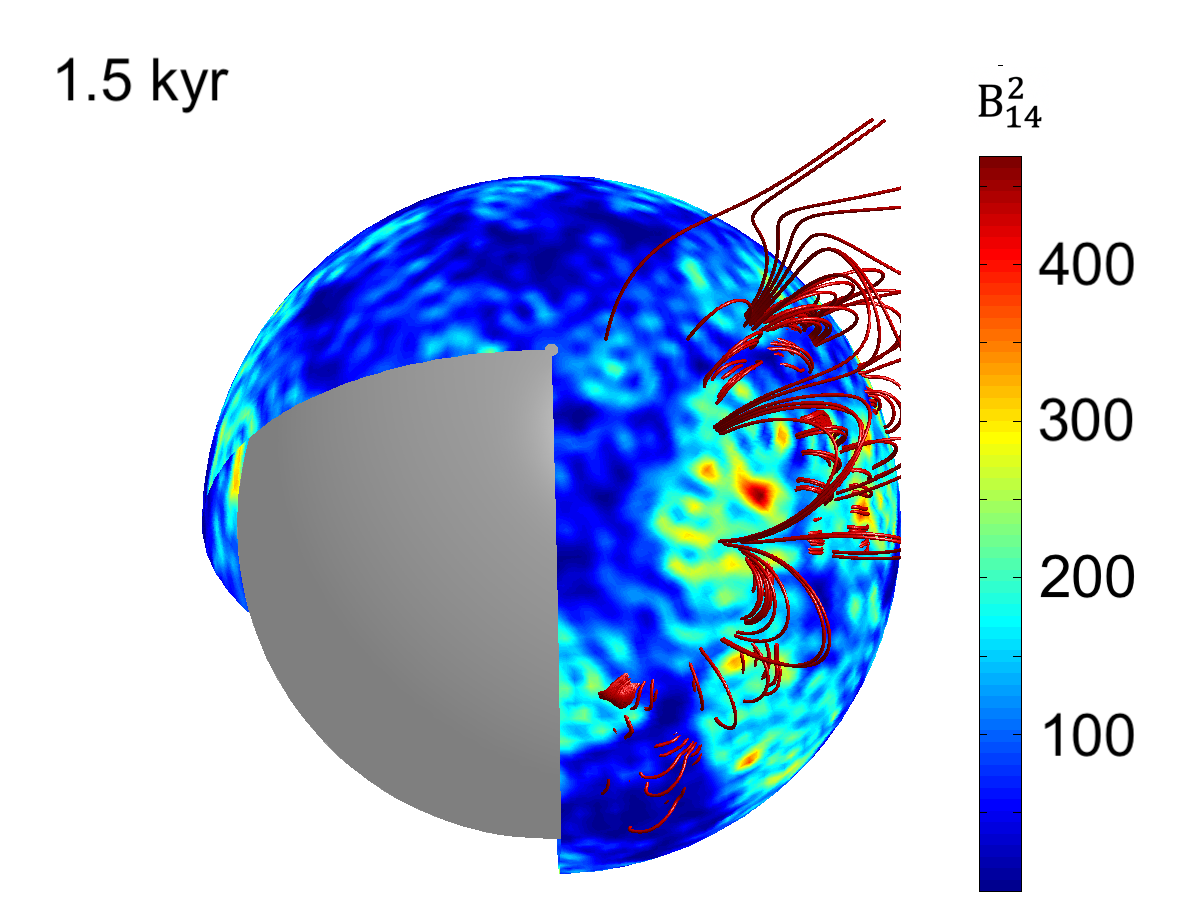}
{\bf c}\includegraphics[width=3.6cm]{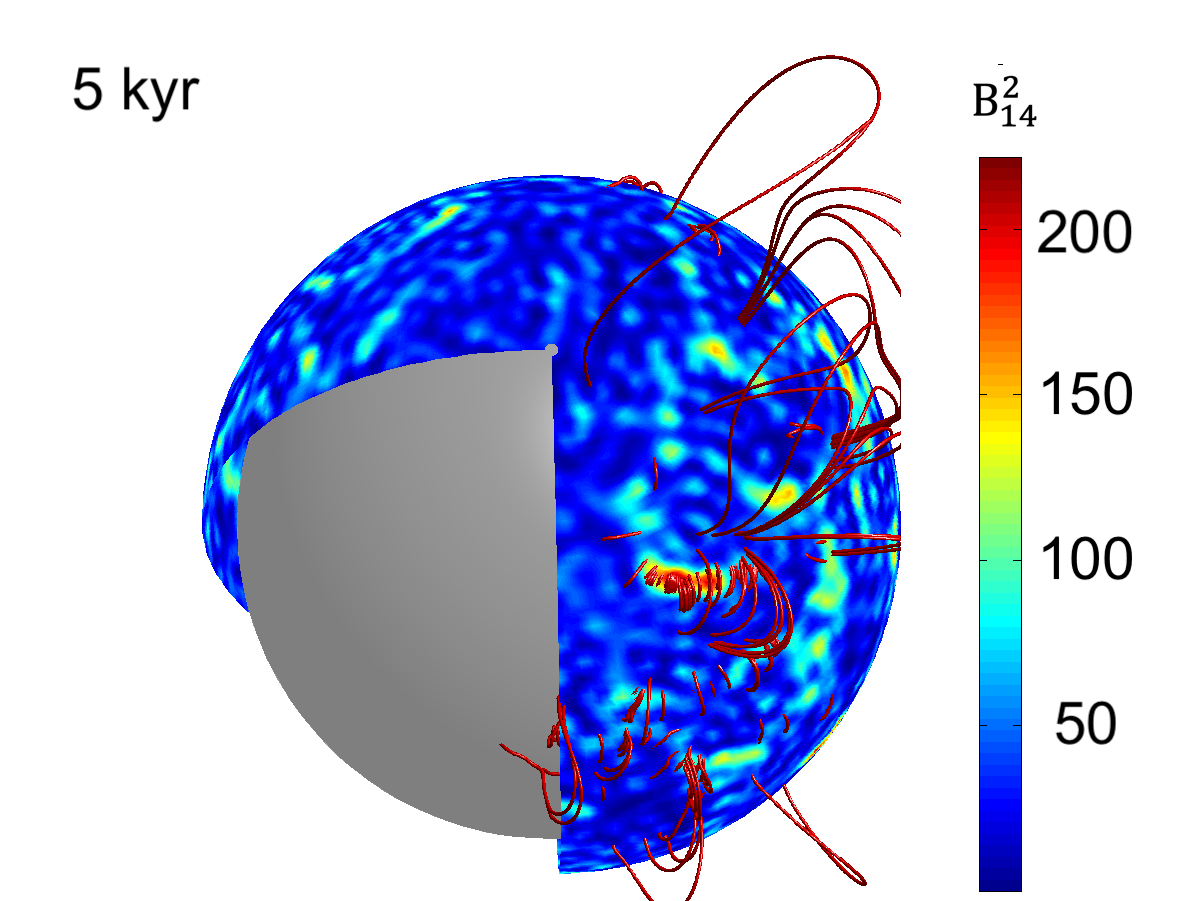}
{\bf d}\includegraphics[width=3.6cm]{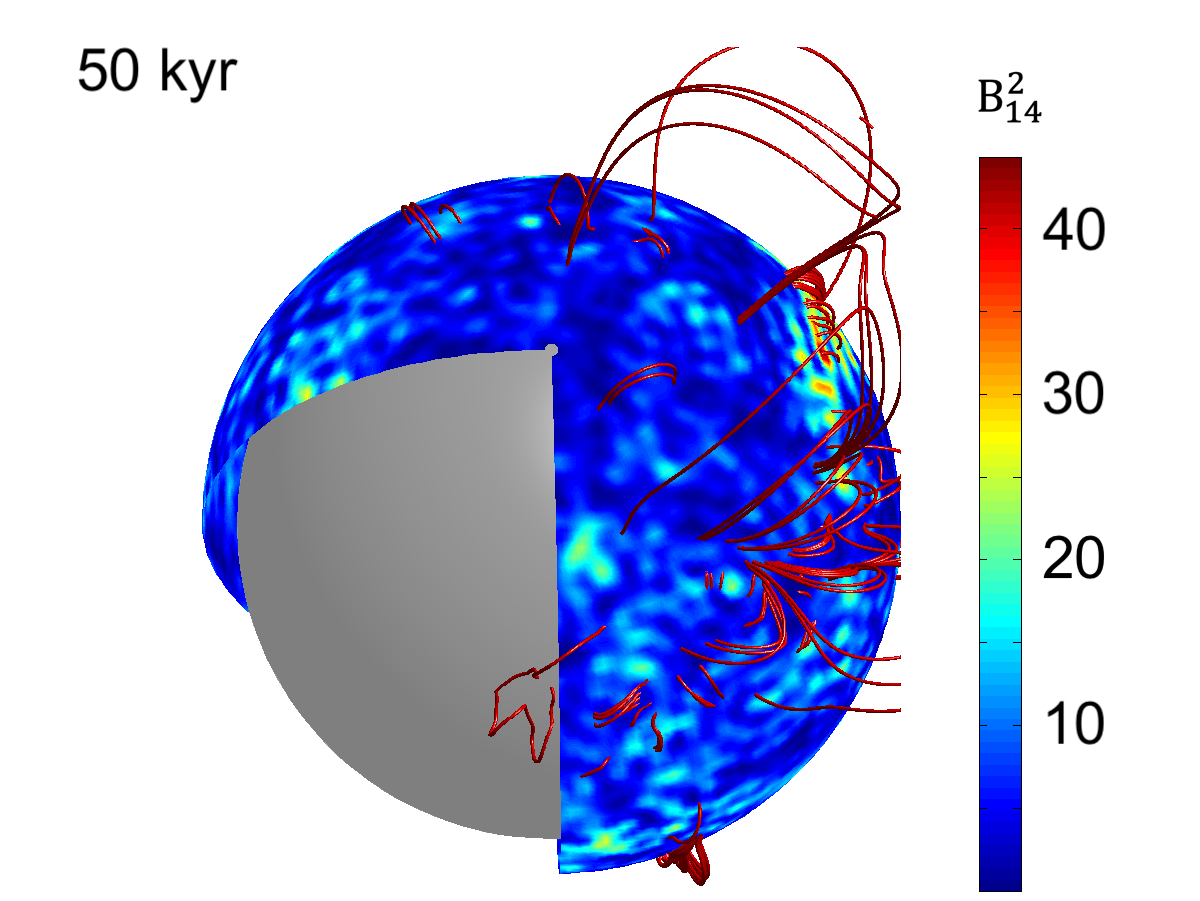}
\caption{The magnetic field structure for Model 6, as in Fig.~\ref{Fig:1}.}
\label{Fig:3}
\end{figure*}

The emerging field has a complex structure on the surface of the star. The maximum strength of the emerging field is a few times smaller than the average initial field, as seen in Figures \ref{Fig:1}-\ref{Fig:3}. The field maintains this strength for a few tens of kyr in models 1-5 and several kyrs for models 5 and 6. At later times the magnetic field decays, but even after 1.5 Myr in models 4 and 5 the complex surface structure remains at a much weaker magnetic field strength: $10^{12}$G compared to the initial $4\times 10^{14}$G. In addition to the complex internal magnetic field, the external part forms arcades over the regions of the surface where the field is the strongest.

\begin{figure*}
\centering
{\bf a}\includegraphics[width=7cm]{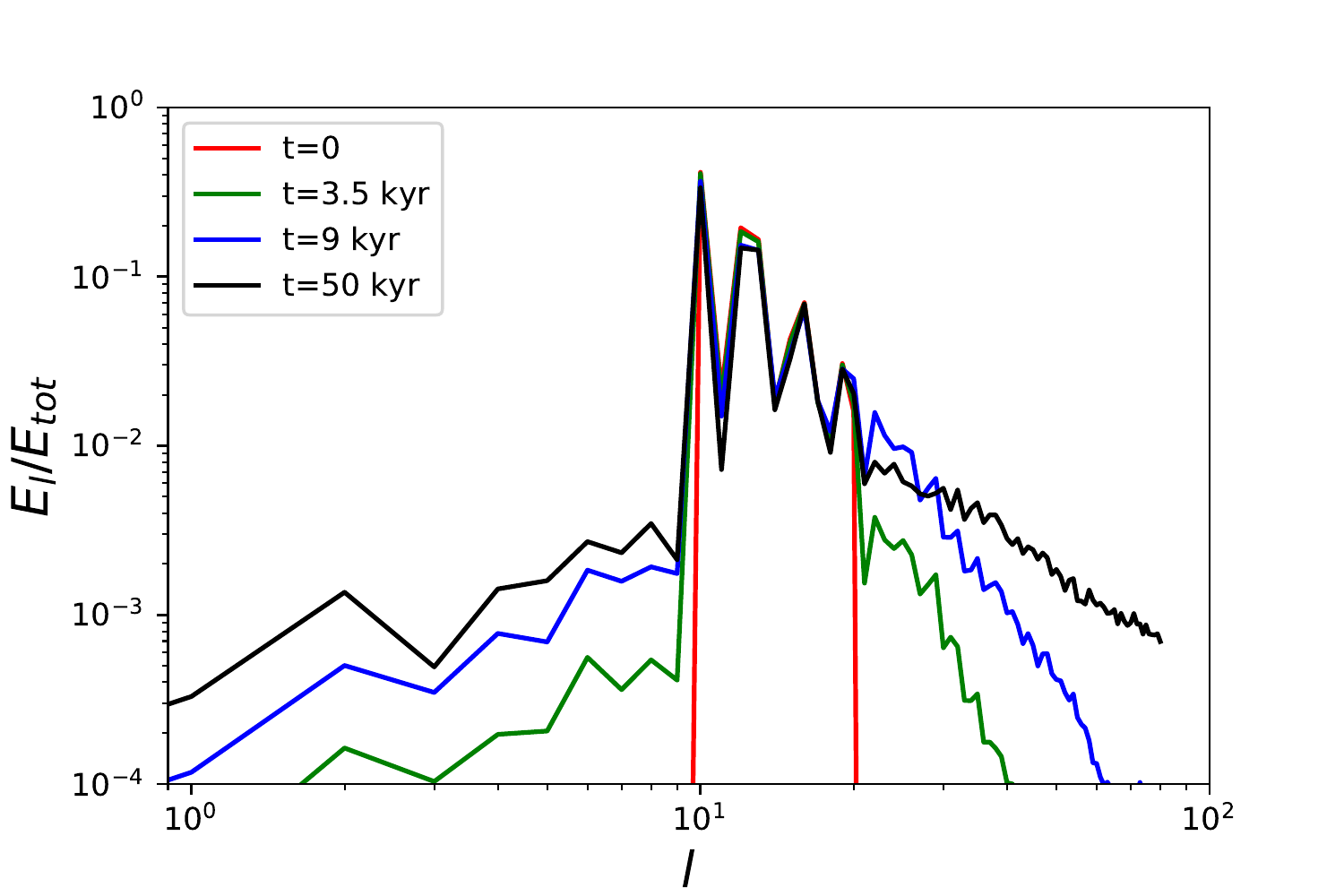}
{\bf b}\includegraphics[width=7cm]{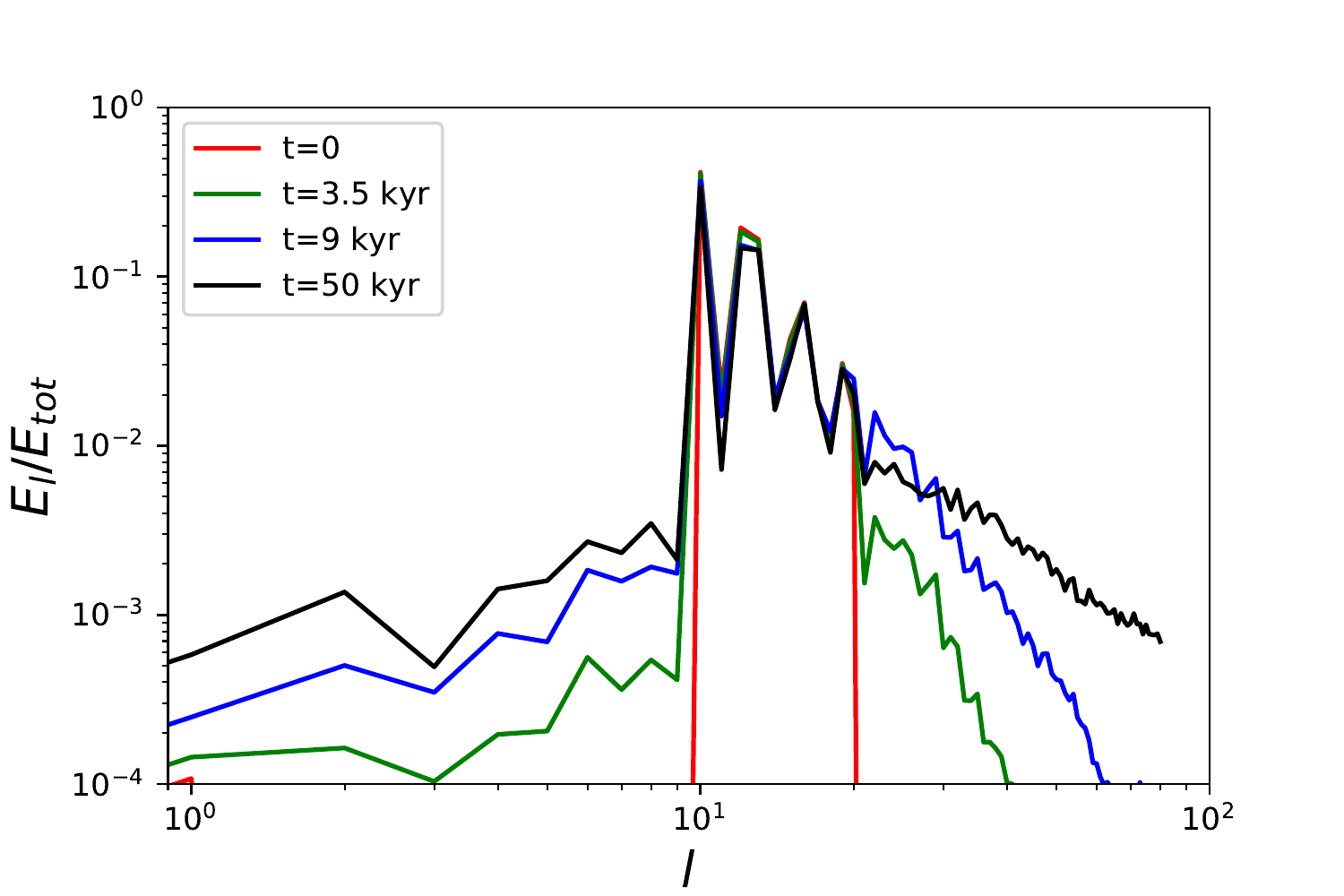}
{\bf c}\includegraphics[width=7cm]{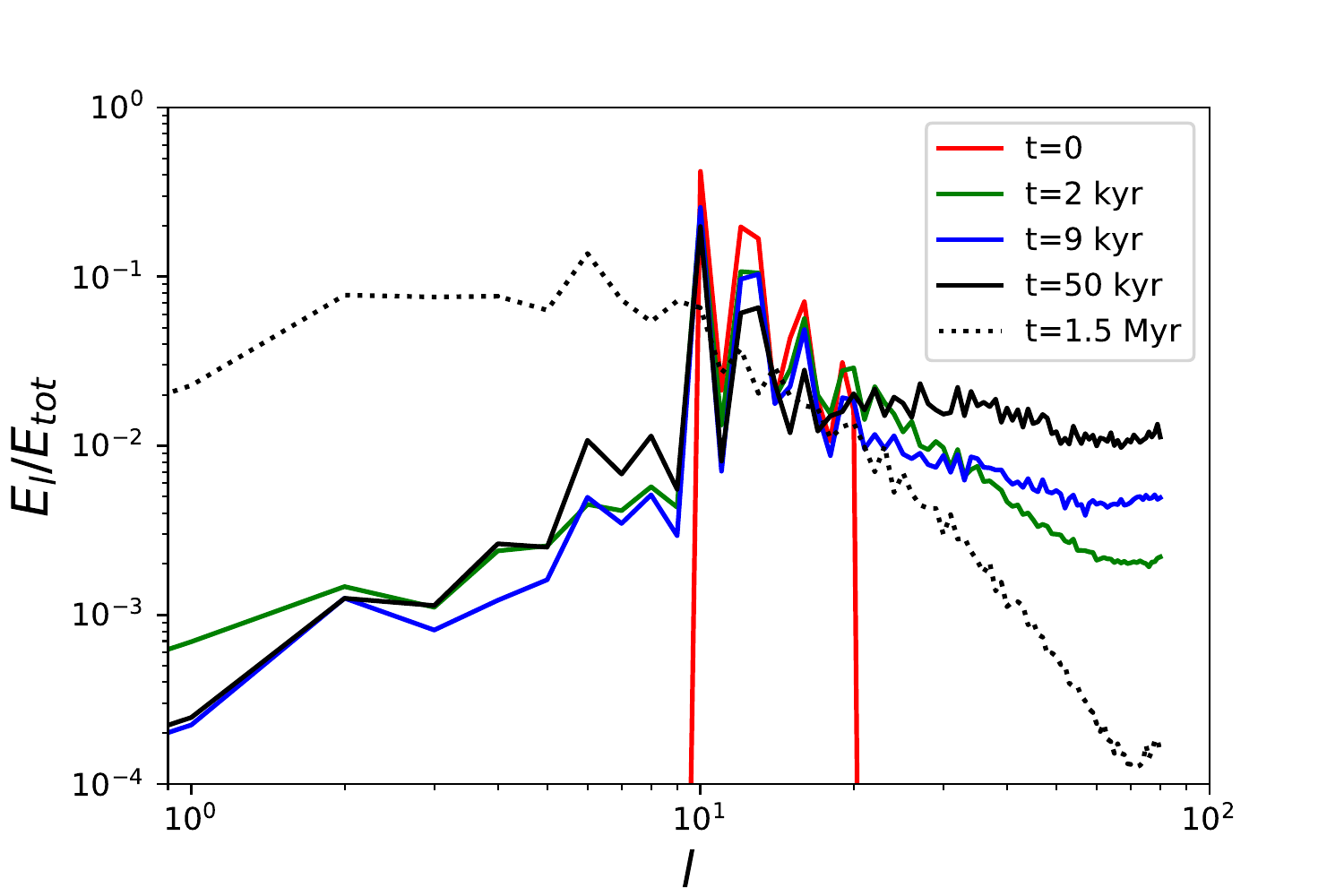}
{\bf d}\includegraphics[width=7cm]{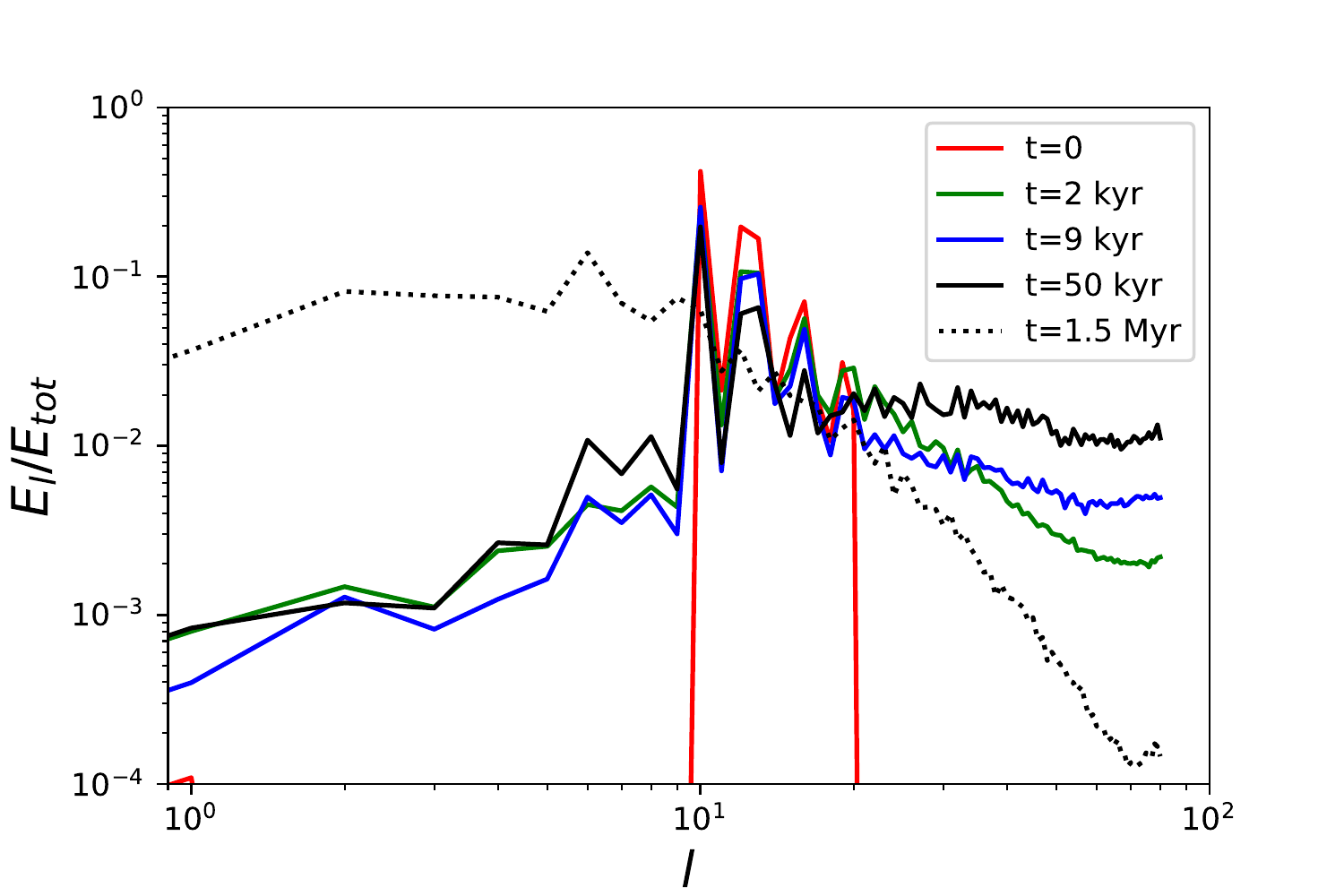}
{\bf e}\includegraphics[width=7cm]{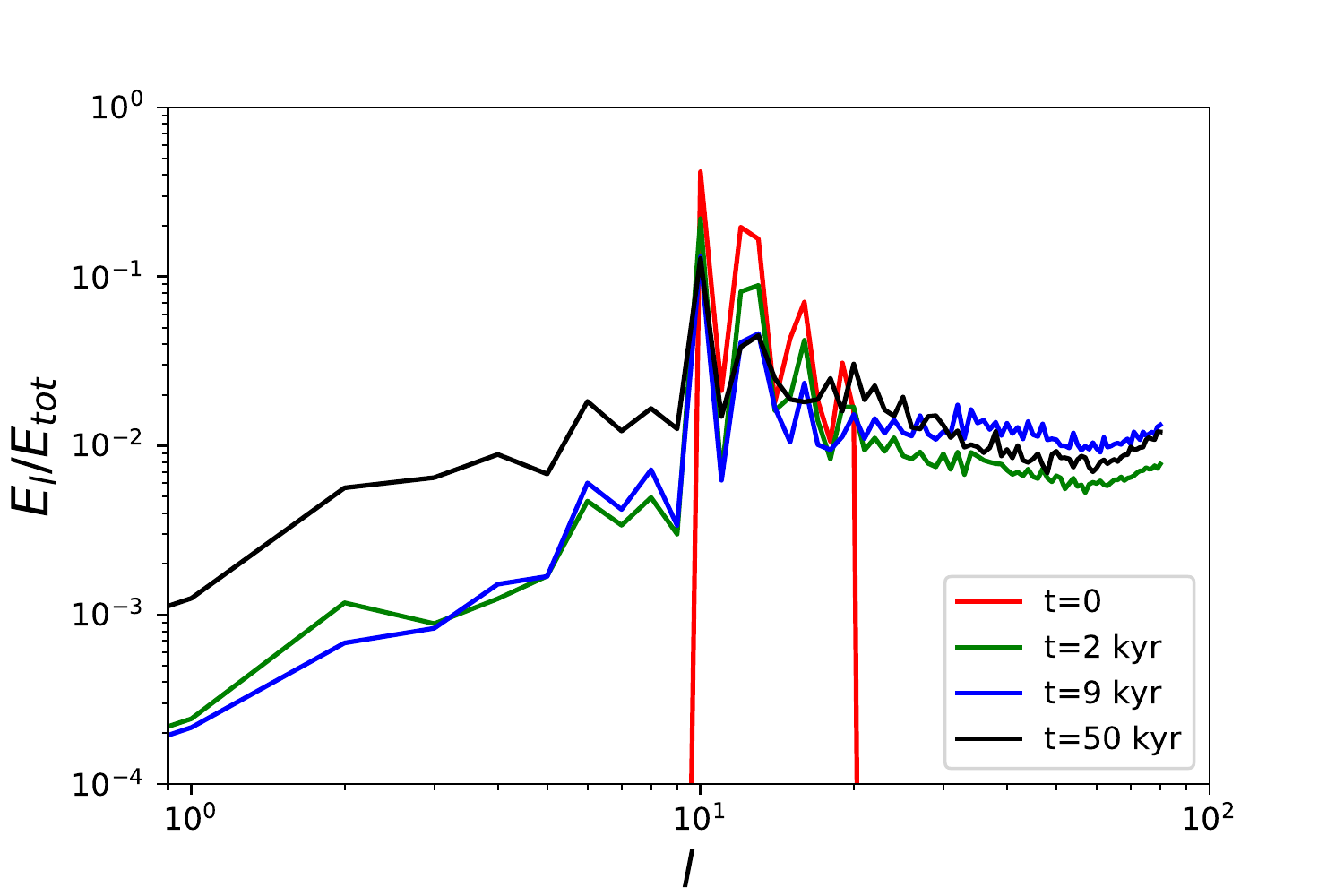}
{\bf f}\includegraphics[width=7cm]{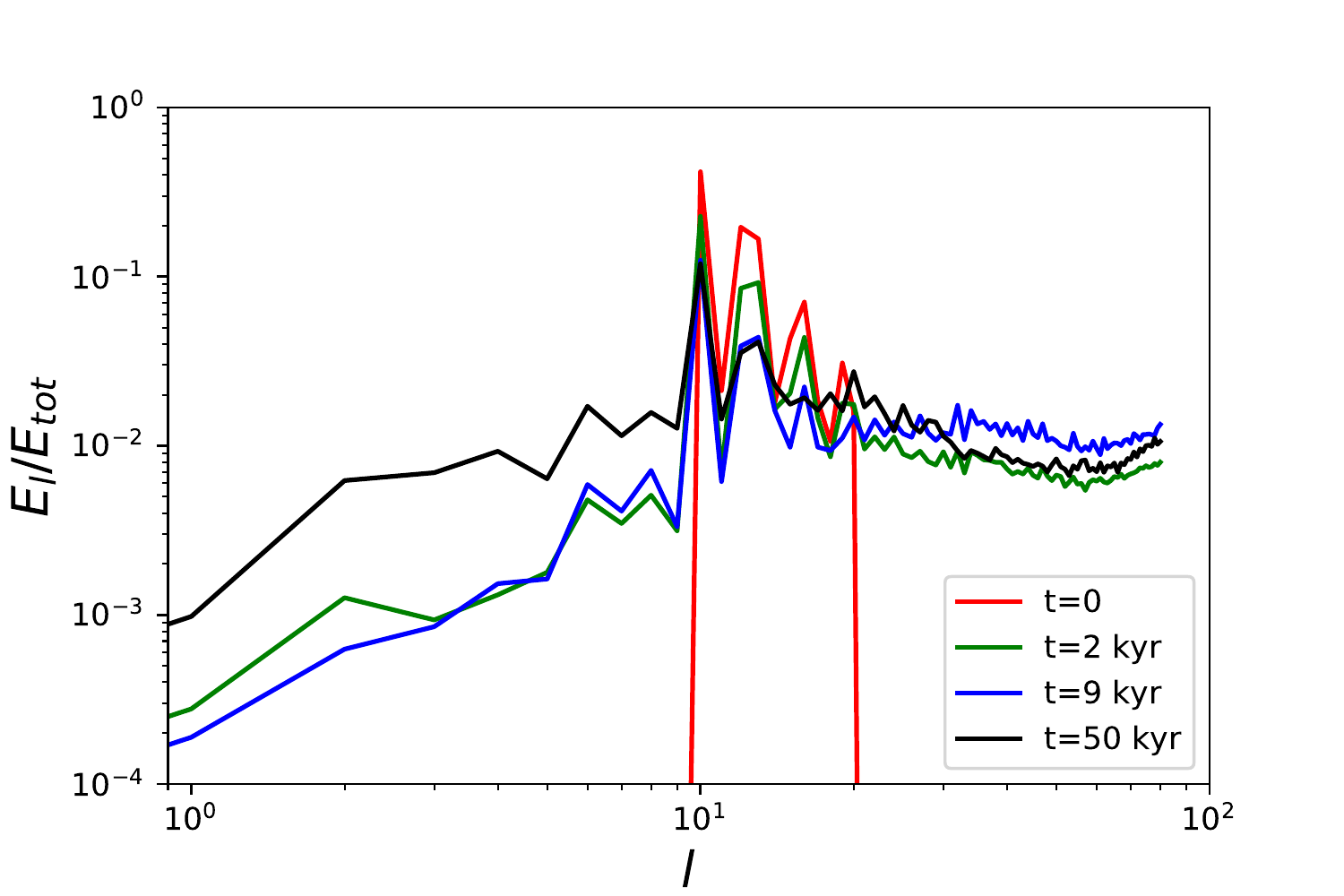}
\caption{The energy in the various $\ell$ multipoles for models 1, 3, 4, 5, 6 and 7, panels (\textbf{a}), (\textbf{b}), (\textbf{c}),  (\textbf{d}), (\textbf{e}) and (\textbf{f}) respectively. }
\label{Fig:4}
\end{figure*}

Seen in spectral space, the magnetic field also populates lower and higher multipoles going down to the dipolar component ($\ell=1$), Figure \ref{Fig:4}. In models 1, 2 and 3 there is a progressive increase in the amount of energy at the lower and higher multipoles at 3.5, 9 and 50 kyrs, suggesting that the dispersion of energy in spectral space evolves until that time.  Note that we plot models 1 and 3 as models 1 and 2 look almost identical. In contrast, in models 4 and 5 the spectral distribution at low $\ell$ is similar at 2, 9 and 50 kyrs, suggesting that the evolution has found some equilibrium. At the very late time $1.5$ Myr the energy at the lower energies clearly dominates. This seems to suggest that the Hall effect pushes energy to both higher and lower parts of the spectrum, but the time it takes to reach the higher multipoles and generate fine structure is longer than the self-organisation and the formation of lower multipoles. The excess of the lower $\ell$'s at very late times is related to the domination of the Ohmic decay, which does not transfer energy between the various multipoles, but leads to the decay of the field, with the higher multipoles decaying faster as they correspond to smaller spatial scale.  In models 6 and 7 where the magnetic field is the strongest, we notice that at 2 and 9 kyrs the spectral distribution at lower and higher $\ell$ is quite similar, suggesting the short time it takes for the system to populate the spectral space and reach a steady-state, once scaled for the global decay. However at 50 kyr, the fraction of the energy at lower multipoles is higher than what it was earlier, whereas the relative amount of energy at higher multipoles is lower there. This suggests that this model has entered the phase where Ohmic decay becomes important to the evolution. This is consistent with the fact that the transfer of energy between different scales is mediated by the Hall effect. As the Hall timescale scales inversely proportional with the magnetic field strength, we expect that this process will be completed faster when the magnetic field is stronger.  We note that as the magnetic field evolves, the peak of the power spectrum remains close to $\ell=10$ during most of the evolution. This is not an intrinsic property of the magnetic field evolution, but instead results from the choice of the initial condition, populating the particular range of multipoles.

The presence of an initial dipole has a mild impact on the later evolution. Comparing models 4 and 5 where the initial dipole component differs by two orders of magnitude, we notice a higher fraction of energy in smaller $\ell$'s by a factor of a few in the dipole component at later times. 

\begin{figure*}
\centering
{\bf a}\includegraphics[width=7cm]{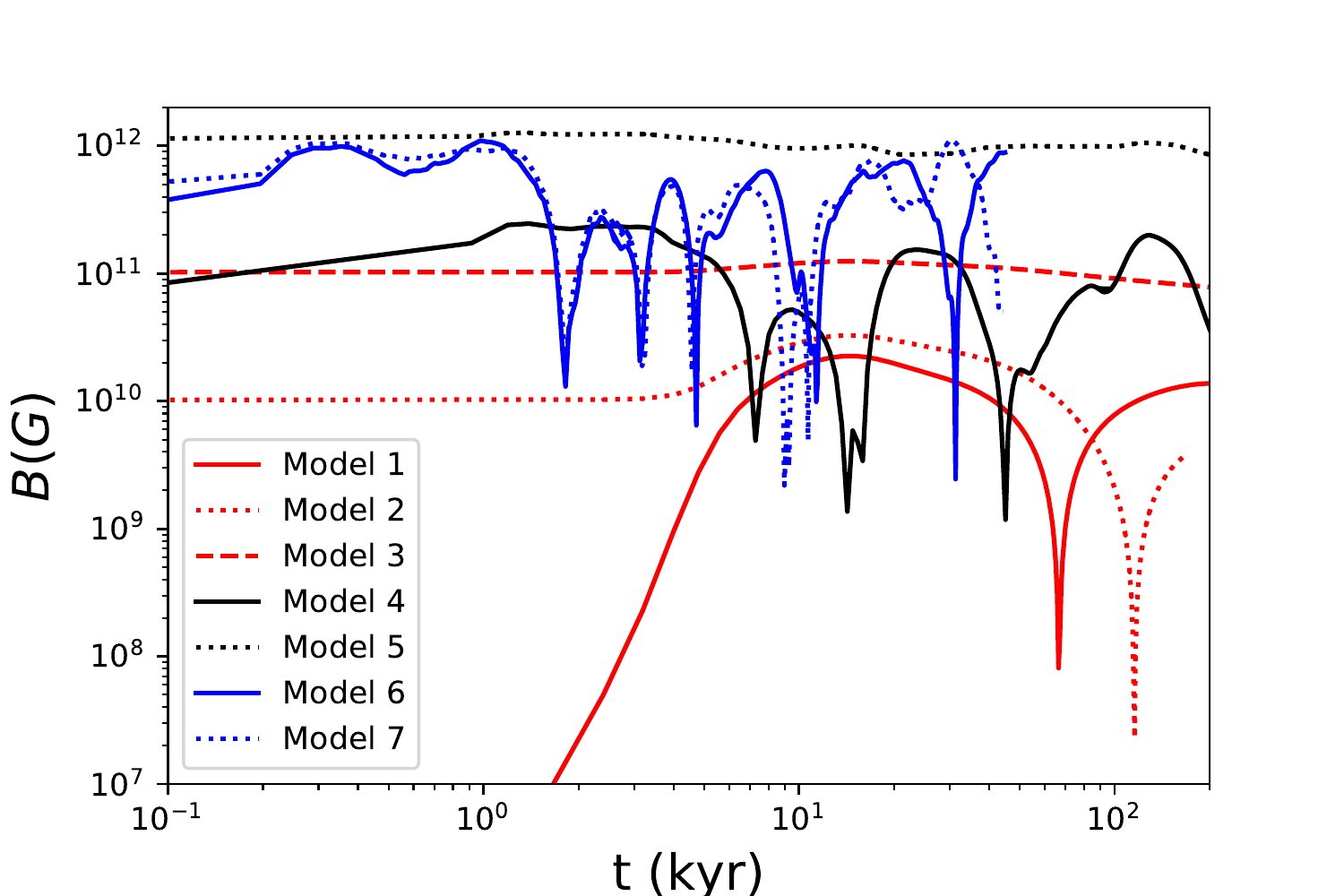}
{\bf b}\includegraphics[width=7cm]{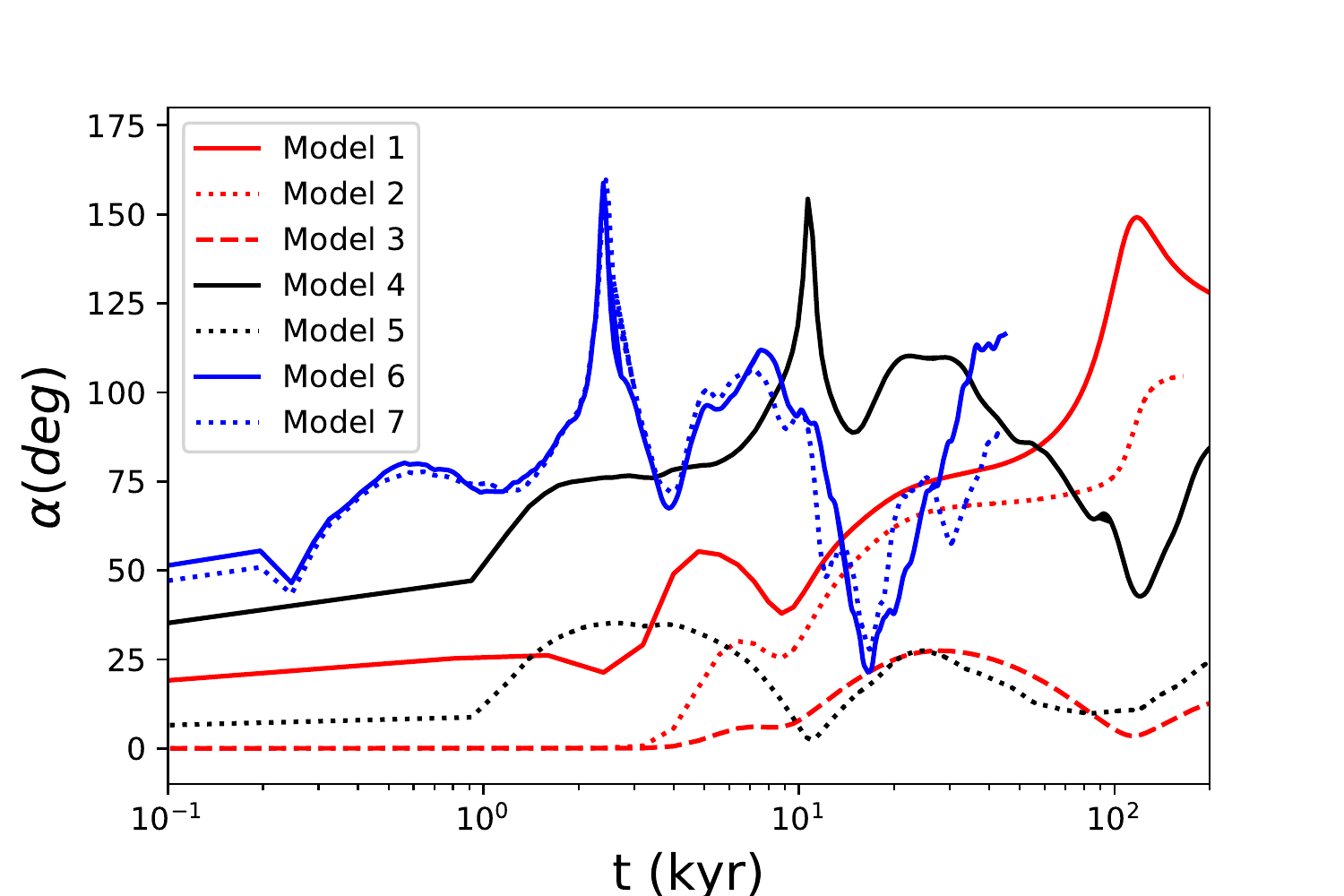}
\caption{Panel {\bf a}: The dipole component of the magnetic field for various models. Panel {\bf b}: The angle of the dipole component with respect to a reference axis. The reference axis is set to the dipole axis at $t=0$ if an initial dipole field is present, otherwise it has a random direction.}
\label{Fig:5}
\end{figure*}

The dipole component of the magnetic field grows to pulsar levels, as shown in the left panel of Figure \ref{Fig:5}, even in the absence of an initial dipole as in model 1. There, a dipole spontaneously forms, at a strength of approximately $10^{10}$ G, about four orders of magnitude less than the internal field. Similarly, in models 4 and 6 the dipole field reaches $2\times 10^{11}$ G and $10^{12}$ G, four orders of magnitude less than the average strength in the interior. Considering model 2, where a dipole field is already present at the initial state, the evolution of this field is similar to the superposition of the initial dipole and the generated field. The long term evolution, especially in the strong field regime (models 6 and 7) is rapid, with the dipole field vanishing for short times and reversing direction. Considering the evolution of the angle of the dipole axis with a reference axis aligned to the direction of the dipole field at $t=0$, we notice that in models 2, 4, 6 and 7 the dipole field reverses direction, which also happens in model 1 which lacks an initial dipole, Figure \ref{Fig:5} right panel. Models 3 and 5, which have a dipole field whose strength is greater than $10^{-4}$ times the average internal field do not have such reversals, as the dipole field is dominated by the initial one.

\begin{figure}
\centering
\includegraphics[width=\columnwidth]{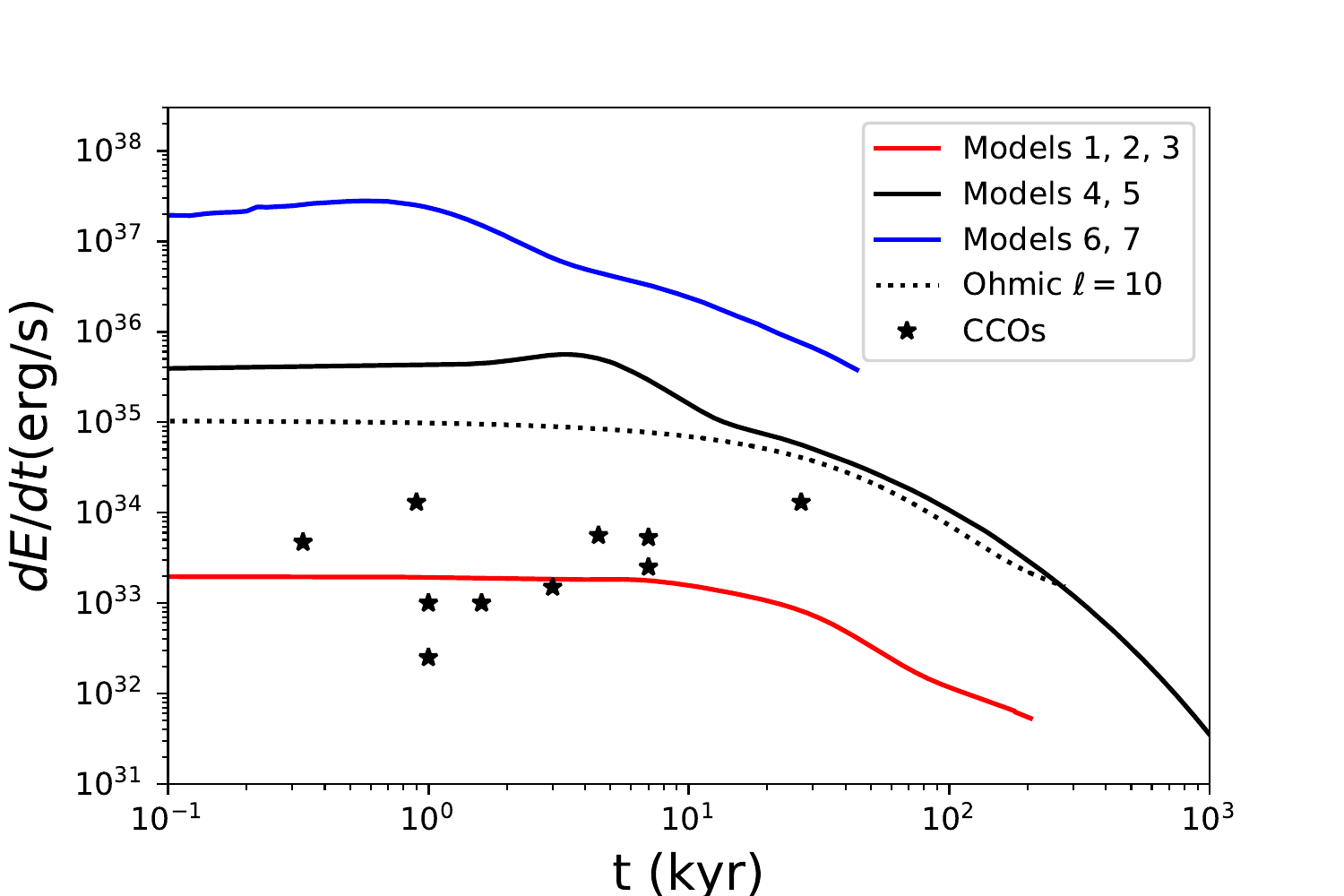}
\caption{The magnetic energy decay rate for various models. The black stars are the CCO confirmed and candidate sources from Table \ref{TAB:2}. The dashed line corresponds to the decay of a multipole with $\ell=10$, $m=0$, with the same radial structure as the models presented here. }
\label{Fig:7}
\end{figure}

The Ohmic decay rate of the magnetic field is depicted in Figure \ref{Fig:7}. As expected the field decays faster in systems where the total magnetic energy is higher, with a scaling steeper than $\bar{B}^2$ especially between the models with the strongest magnetic field. In the families of models with the strongest field (4, 5) and (6, 7) the decay rate becomes maximum at a slightly later time, 5 and 1 kyr respectively. This is mainly because the Hall effect requires some time to generate smaller structures that will lead to faster magnetic field decay. Compared to the pure Ohmic decay of an $\ell=10,\, m=0$ multipole, we notice that the decay rate is decreasing monotonically and is overall slower. 

\section{Discussion}
\label{DISCUSSION}

The outstandingly weak dipole field of the three CCOs where timing measurements were possible has led to the term ``anti-magnetar" \cite{Halpern:2010}. CCOs have a puzzling behaviour: their dipole inferred magnetic field is rather weak \citep{Halpern:2007, Gotthelf:2007}, while their X-ray luminosity in some sources exceeds $10^{33}$erg~s$^{-1}$, which overlaps with that of magnetars. Unlike magnetars, however, CCOs do not have bursting or flaring behaviour, implying that the mechanisms that could power magnetar bursts, such as crust yielding, are not operating in these sources. The X-ray spectrum of these sources consists of one or two blackbody components \citep{Gotthelf:2010}, which leaves no room for inverse Compton scattering typically seen in magnetars due to magnetospheric twist.

Magnetic field decay is a rather efficient mechanism to provide thermal luminosity \citep{Pons:2007, Pons:2009}. However, if an exceptionally strong magnetic field is assumed, this will inevitably lead to crust yielding or magnetospheric instabilities which would be seen as bursts \citep{Perna:2011, Pons:2011}. This hurdle could be overcome if the magnetic field, instead of being exceptionally strong and large-scale, is weaker but has a smaller-scale structure. As the Ohmic decay rate is proportional to electric current squared, a weaker magnetic field with a finer structure will still be supported by a strong current which is what eventually determines the decay rate. 

Among the simulations we have discussed, models 1-3 satisfy this condition: the average field inside the crust is approximately $10^{14}$ G, leading to a Maxwell stress  below the critical limit for the crust to yield \citep{Lander:2015}. Nevertheless, owing to the fact that the field consists of high multipoles, the decay rate is fast enough to power CCOs. Hall evolution and emergence of the magnetic field leads to regions on the surface where the magnetic field is substantially stronger than the average surface value. While a full magneto-thermal calculation is needed to estimate the temperature and luminosity of the sources, these regions can be associated to the pulsed emission from some CCOs. Such regions do exist, but they are not concentrated in the form of a single spot, instead they are spread out in various areas. The pulsed fraction follows the same pattern in almost all CCOs; it is typically low with the exception of Kes 79 being 64\%. Taking into account the structure of the external field, we can qualitatively attribute absorption features to the arcades forming above the regions of strong surface magnetic field. Models with stronger initial magnetic field (4-7) have strengths which could potentially make the crust yield and dipole fields that are around or slightly above $10^{11}$G. Such values are stronger than what is observed in CCOs.   

\begin{figure}
\centering
{\bf a}\includegraphics[width=\columnwidth]{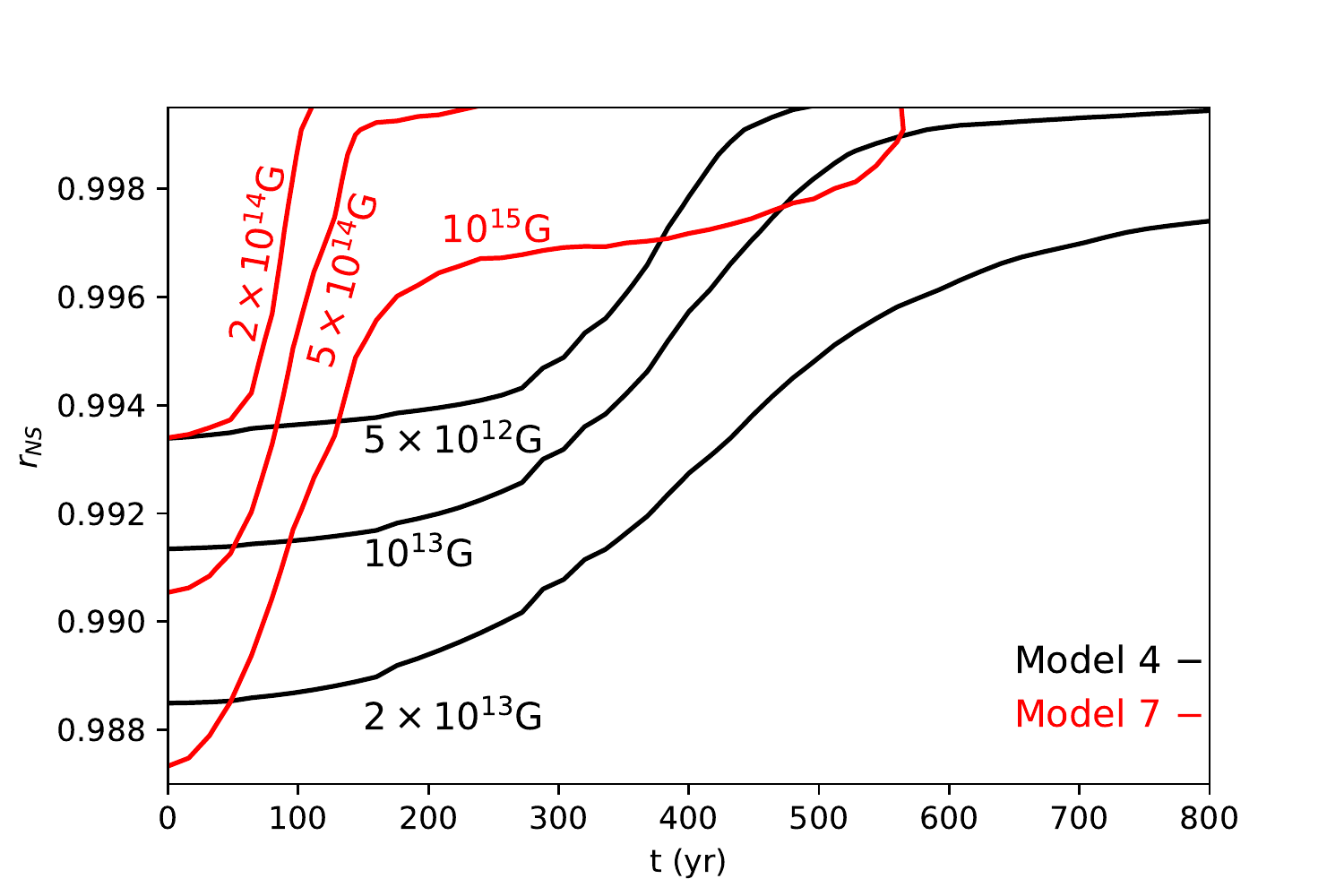}
\caption{The emergence of the magnetic field to the neutron star surface for models 4 and 7. We plot the shell averaged value of the magnetic field as it moves towards the surface of the star. The black lines correspond to model 4 and the red to model 7.}
\label{Fig:Emergence}
\end{figure}

 The emergence of the field at the surface is very fast in the strong magnetic field models,  taking about 100 years for models 6 and 7. This timescale becomes 500 years for the moderately magnetised models 4 and 5. In the weakly magnetised models (1-3) it can exceed 2000 years, with the field essentially just diffusing ohmically towards the surface. For the quantification of the emergence time, and taking into account the highly anisotropic structure of the field, we have averaged the magnetic field at every shell and we consider the time it takes for the magnetic field at the surface to reach the value the field had at $0.94r_{NS}$ at $t=0$. In Fig.~\ref{Fig:Emergence} we plot the shell-averaged value of the magnetic field as it emerges to the surface of the star for models 4 and 7. We note that the behaviour of models 5 and 6 is rather similar to models 4 and 7 respectively. As the field is highly anisotropic, regions of much stronger magnetic field appear on the surface at about half the time it takes for the average field on the surface to increase. The emergence of a buried field is in qualitative agreement with past work \citep{Geppert:1999, Ho:2011, Vigano:2012, Shabaltas:2012, Igoshev:2016}. The main step forward here is that the field is not axisymmetric but rather has a complex fully 3-D geometry. Thanks to this geometry it is possible to have an overall faster evolution, due to the fact that currents are higher and this is what determines the electron fluid velocity and eventually the advection of the magnetic field.  

The spontaneous formation of a dipole magnetic field is caused by the nonlinear nature of the Hall effect. Ohmic evolution alone would be linear, leading to an exponential decay of the various modes without energy exchange between different spherical harmonics. Here, the magnetic field instead gets redistributed across spectral space, occupying both higher and lower $\ell$'s. However, no signs of a true inverse cascade are present, as the peak stays within the initially excited modes. This is in agreement with previous Hall simulations that have found that the Hall effect leads to an inverse cascade in 2-D geometries but  not in 3-D \citep{Wareing:2009b, Wareing:2010}; \cite{Igoshev:2016} also found that the initially excited modes stay present for a long period of evolution even in 2-D geometry.

\begin{figure*}
\centering
{\bf a}\includegraphics[width=7cm]{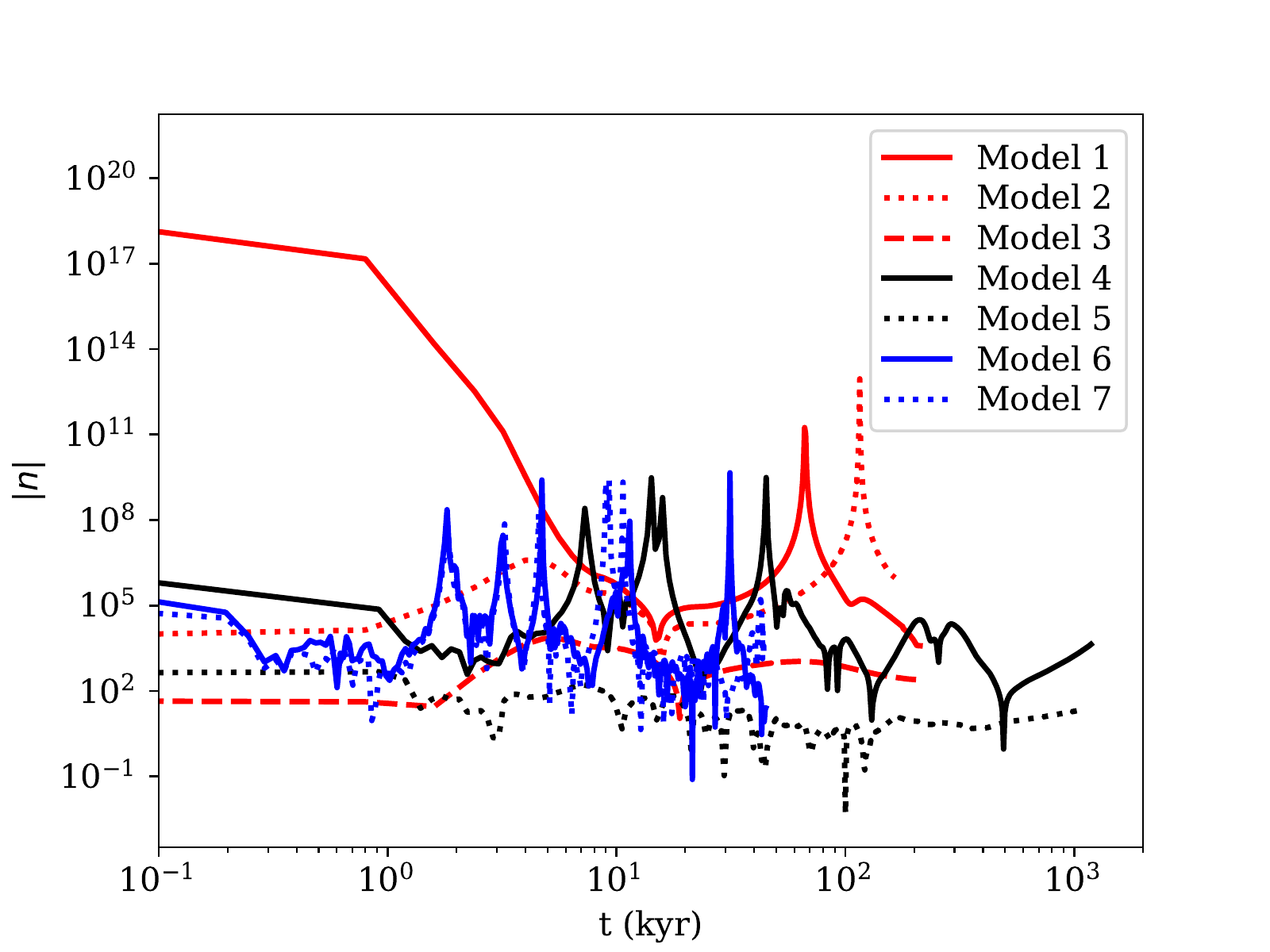}
{\bf b}\includegraphics[width=7cm]{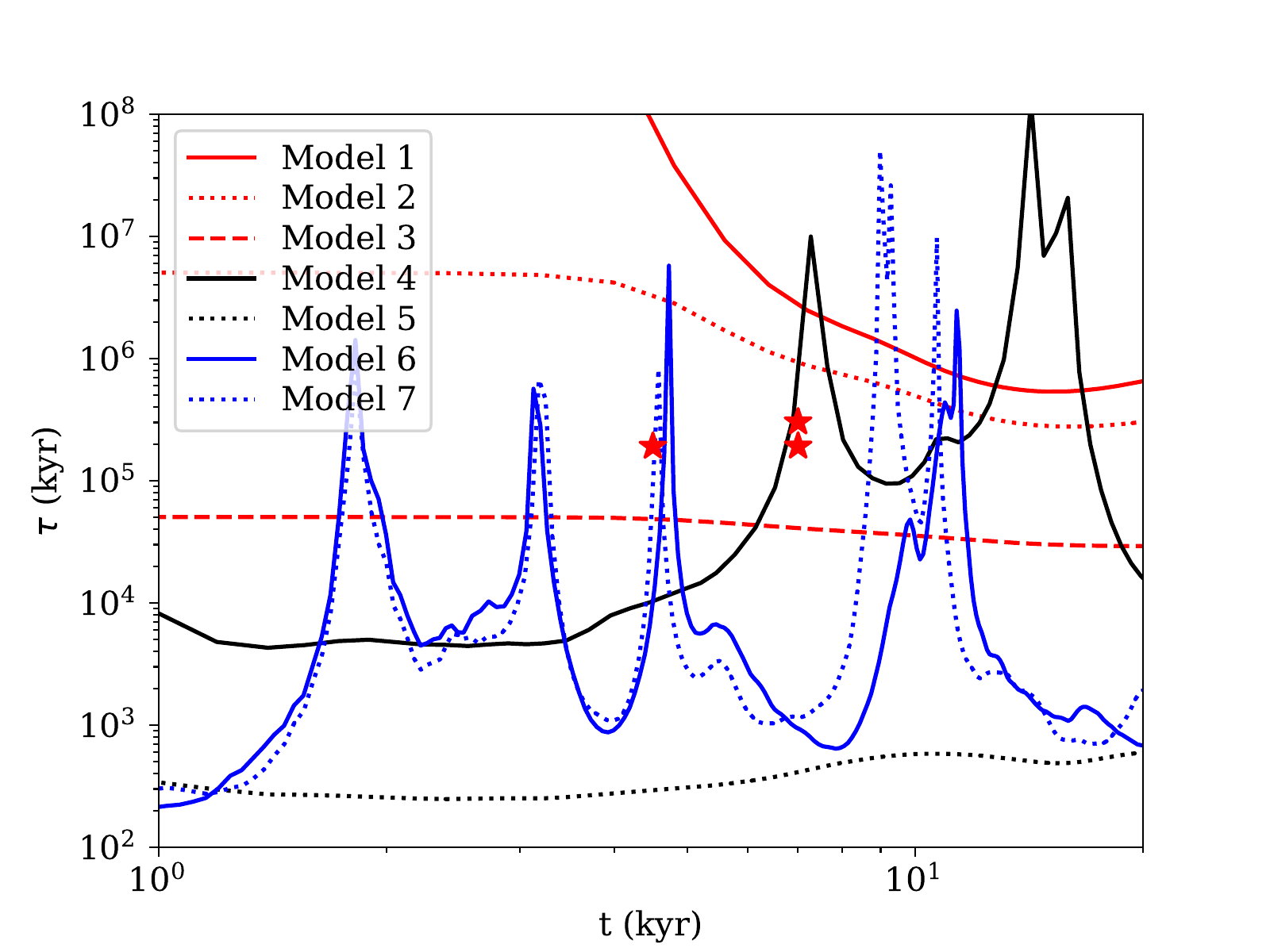}
\caption{Panel {\bf a}: The evolution of braking index for various models. Panel {\bf b}: the evolution of spin-down age $\tau$ as a function of actual age for various models. Red asterisks show location of the three CCOs with measured period and period derivatives. Actual ages for these CCOs are estimated based on the associated supernova remnant age.}
\label{Fig:9}
\end{figure*}

The evolution of such neutron stars in the $P-\dot{P}$ diagram is in essence a vertical oscillation parallel to the $\dot{P}$ axis, except for model 5 whose initial state contains a strong dipole field. This is because their magnetic field evolves drastically with time, while remaining weak enough that it cannot spin them down noticeably. Thus, their period changes very mildly, even if it was in the order of $0.1$ s at birth. The fact that the dipole strength changes a lot suggests that the sources may move in and out of the more densely populated areas of the $P-\dot{P}$ diagram. These oscillations in the  $P-\dot{P}$ diagram translate into oscillations of the spin-down age, see Figure~\ref{Fig:9}. Overall, the spin-down ages are a few orders of magnitude larger in our simulations in comparison to actual ages. We predict that these spin-down ages either stay nearly constant as in the case of models 2, 3, 5 or goes up and down as in models 4, 6, 7. The spin-down ages start growing steadily only when the actual age reaches $0.1-1$~Gyr.

\begin{figure}
\centering
\includegraphics[width=\columnwidth]{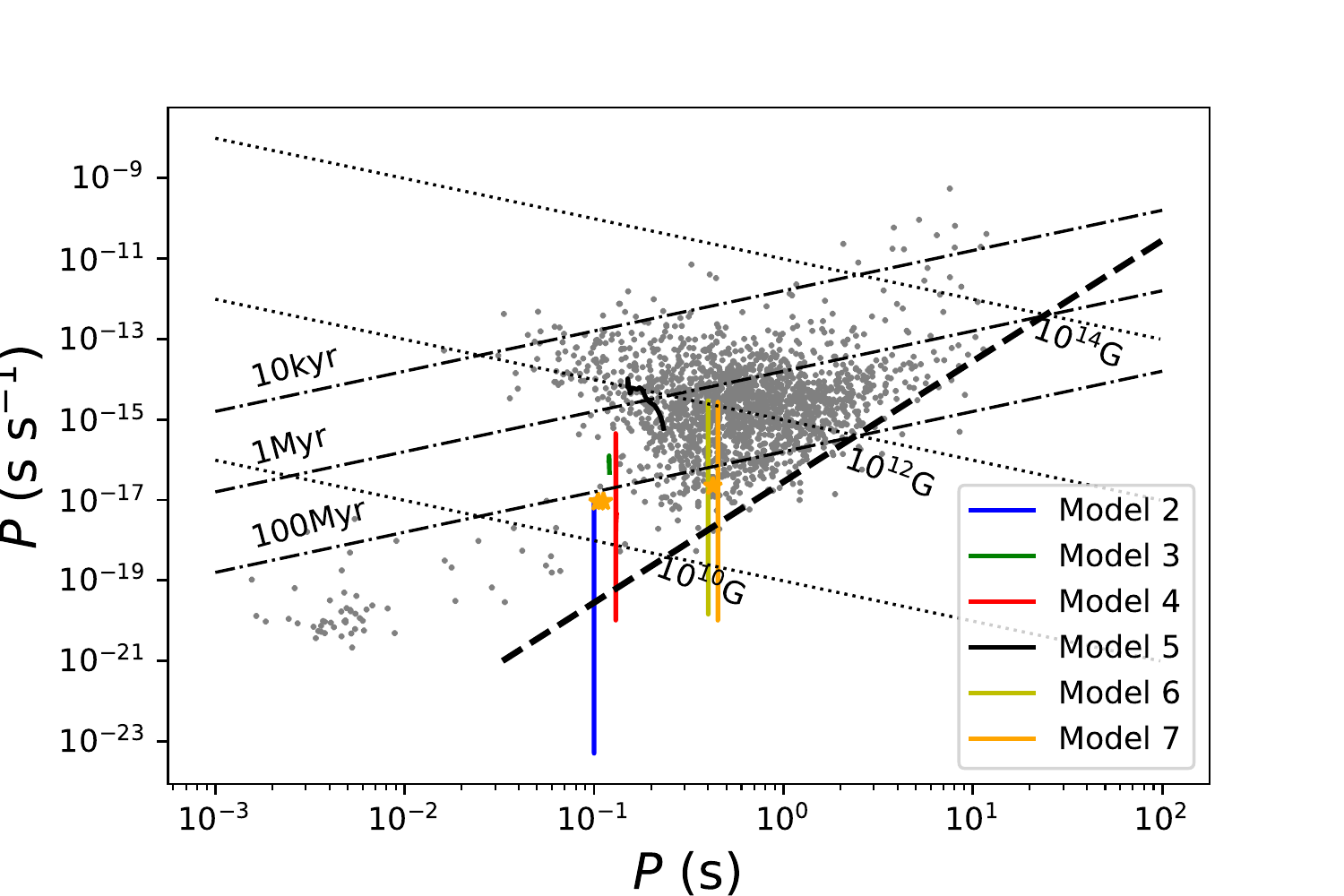}
\caption{The evolution of pulsars in the $P-\dot{P}$ diagram for various models. In models 2-5 the value of the initial period is set between $100$-$150$ ms; in models 6 and 7 the initial period is set to $420$-$450$ ms.  Note, that as model 1 starts with a dipole field identically zero its track will follow that of model 2 with a delay of about 1 kyr.}
\label{Fig:8}
\end{figure}

If any of these neutron stars develop conditions allowing copious pair formation and start operating as radio pulsars, the properties of such a pulsar will be quite remarkable. The braking index $n=2-P\ddot P/\dot P^2$ is expected to be far away from its classical value of 3 typical for magnetic dipole braking in vacuum, see Figure~\ref{Fig:9} for absolute values of $n$. We compute this plot using the equation for braking derived in \cite{Philippov:2013}. The reasons for extremely large braking index is the constant change of strength and orientation of the dipolar magnetic moment. For most of our models, the braking index is around $10^3$, with the exception of the model 5 which behaves quite similar to a normal pulsar, see the track in the $P-\dot P$ in Figure~\ref{Fig:8}. Such braking indexes could be seen as so-called `red' noise in the pulsar timing if no care is taken to consider high-order derivatives as well. Additional peculiarity of these pulsars would be their long-term intermittency. These pulsars could experience multiple episodes of turning on and off on timescales of kyrs, in addition to becoming invisible due to the shift of magnetic pole with respect to the observer's line of sight.

The surface magnetic field of model 4 stays at a quite constant value of a few $10^{10}$~G for a Myr. Neutron stars with similar magnetic fields are detected in radio. Moreover, a wide range of surface $\ell$ harmonics could create a small curvature radius of open field lines, and so facilitate copious pair formation and pulsar activation \citep{Igoshev:2016}. Recent numerical simulations \citep{Philippov:2020} also confirmed an essential role of the field line curvature for producing pulsar radio emission. Previous searches for descendants of post-CCO pulsars did not find any reliable candidates \citep{Bogdanov:2014, Luo:2015,Igoshev:2018}. These searches aimed at pulsars with large spin-down ages and excessive bulk surface X-ray emission. However, at ages of Myr when the magnetic field stabilises enough, the bulk surface temperature goes well beyond the current observational limit for soft X-rays. Therefore, a study of braking indexes for old radio pulsars might be the only viable way to discover CCO descendants.

\begin{table*}
\centering
\caption{Central Compact Objects, data adapted from \protect\cite{DeLuca:2017} \href{http://www.iasf-milano.inaf.it/~deluca/cco/main.htm}{http://www.iasf-milano.inaf.it/$\sim$deluca/cco/main.htm}.}
\begin{tabular}{cccccccc}
\hline
CCO &	SNR	& SNR Age&	d&	P	&PF&	$B_s$&	$L_x$ bol\\
		 & &(kyr)&	(kpc)	&(s)&	(\%)	&(10$^{10}$ G)&	(erg s$^{-1}$)\\
		\hline
RX J0822.0-4300	&Puppis A	&4.5	&2.2	&0.112	&11&	2.9&	$5.6\times 10^{33}$ \\
CXOU J085201.4-461753&	G266.1-1.2&	1	&1&	-	&$<7$&	-&	$2.5 \times 10^{32}$\\
1E 1207.4-5209	&PKS 1209-51/52&	7&	2.2&	0.424	&9	&9.8&	$2.5 \times  10^{33}$\\
CXOU J160103.1-513353	&G330.2+1.0	& $>3$&	5	&-&	$<40$&	-&	$1.5 \times  10^{33}$\\
1WGA J1713.4-3949&	G347.3-0.5&	1.6&	1.3&	-&	$<7$	&-	&$\sim 1 \times  10^{33}$\\
XMMU J172054.5-372652&	G350.1-0.3	&0.9&	4.5	&-&	-&	-&	$3.9 \times  10^{33}$\\
XMMU J173203.3-344518	&G353.6-0.7&	$\sim$27	&3.2&	-&	$<8$&	-&	$1.3 \times  10^{34}$\\
CXOU J181852.0-150213&	G15.9+0.2	&1--3	&(8.5)&	-&	-&	-&	$\sim 1 \times  10^{33}$\\
CXOU J185238.6+004020&	Kes 79	&7	&7	&0.105&	64&	3.1&	$5.3 \times  10^{33}$\\
CXOU J232327.9+584842&	Cas A&	0.33	&3.4&	-&	$<12$&	-&	$4.7 \times  10^{33}$\\
\hline
\end{tabular}
\label{TAB:2}
\end{table*}

\section{Conclusions}
\label{CONCLUSIONS}

Motivated by the peculiar behaviour of CCOs, we have simulated the magnetic field evolution under the Hall effect and Ohmic dissipation for initial conditions where the field has a tangled structure, is buried within the crust, and the dipole component is either completely absent or very weak. We have set the scale of the magnetic loops to be in the order of $\sim 1$~km. Our main conclusions are that:
\begin{enumerate}
\item The magnetic field drifts towards the surface of the neutron star, on a timescale which is shorter for stronger magnetic fields.
\item The surface field has a strength which is a few times less than the internal field, and maintains this highly non-dipolar structure for several tens of kyrs of evolution.
\item The field spontaneously generates a dipole component, whose strength is approximately $10^{-4}$ times the mean strength of the magnetic field inside the crust.
\item While higher and lower multipoles are populated, the peak of the power spectrum does not shift to a lower multipole, suggesting that there is no inverse cascade, as expected from previous 3-D Hall simulations \citep{Wareing:2010}.
\item The decay rate of the magnetic field is fast enough to provide sufficient Ohmic heating to power CCOs, assuming that the internal field is $\sim 10^{14}$~G. Such a field is not strong enough to lead to crustal yielding and bursting behaviour.  
The fast magnetic field evolution occurring in the crust could be seen through anomalous values of braking index $n\sim 10^3$, if any of these objects eventually develop conditions favouring the activation of the radio pulsar mechanism.
\end{enumerate}

We suggest that the presence of a tangled magnetic field as a power source for CCOs is a plausible scenario. This is consistent with the fact that CCOs have $P>0.1$ s, reflecting their period at birth. Dynamo operation in proto-neutron stars with such rotation rates is not efficient, and may not lead to the formation of a strong dipole field \citep{Naso:2008}. Thus, there still exists magnetic flux from convective activity, but it never gets organised into an ordered dipole field. A tangled field gets dissipated faster than a dipole containing the same amount of energy, since the length-scales of the former are much smaller. Therefore a CCO can be powered by such a field, reaching magnetar-level X-ray luminosities without having the accompanying explosive activity. In this sense the term anti-magnetar that has been proposed for CCOs reflects the magnetar X-ray luminosity levels without the need of magnetar magnetic field strengths.
 
\vspace{6pt}



\section*{Acknowledgments}
This work was supported by STFC grants ST/K000853/1 and ST/S000275/1. The numerical simulations were carried out on the STFC-funded DiRAC I UKMHD Science Consortia machine, hosted as part of and enabled through the ARC HPC resources and support team at the University of Leeds. 
K.N.G. acknowledges support from COST ACTION PHAROS (CA16214) for the participation in ``Magnetic field formation and evolution in neutron stars'', the CoCoNuT meeting in Saclay 14-16 November 2018, and the stimulating discussions that inspired this work. K.N.G. thanks Wyn Ho for insightful discussions about CCOs and the physical processes taking place as the magnetic field is buried and re-emerges.
We thank an anonymous referee for insightful comments that helped us to improved this manuscript.  

\newpage
\bibliographystyle{mnras}
\bibliography{BibTex.bib}

\label{lastpage}

\end{document}